\documentclass[showpacs, twocolumn, aps, prl, longbibliography, floatfix]{revtex4-2} %[twocolumn,npa,showpacs]
        \expandafter\let\csname equation*\endcsname\relax
        \expandafter\let\csname endequation*\endcsname\relax
        \usepackage{amsmath,bm}
        \usepackage{multirow,booktabs}   % for tables 
        \usepackage{amsthm}
        \usepackage{multibib}
        \usepackage[english]{babel}
        
        \usepackage{mathrsfs}
        \usepackage{amssymb}
        \usepackage{longtable}   % for very long multipage table 
        \usepackage{subfigure}
        \usepackage{float}

        \usepackage{amsfonts} 
        \usepackage{graphicx}
        \usepackage{url}
        \usepackage{hyperref}
        \usepackage[columnwise]{lineno}  
        \usepackage{color}
        \usepackage{comment} 
        \usepackage{epstopdf}  %when compile to pdf with eps figures, this is needed
        
        \usepackage{tikz}
        \usetikzlibrary{tikzmark}
        \renewcommand{\comment}[1]{}
        \newcommand{\red}[1]{\textcolor{red}{#1}}

            \makeatletter
            \newsavebox{\@brx}
            \newcommand{\llangle}[1][]{\savebox{\@brx}{\(\m@th{#1\langle}\)}%
              \mathopen{\copy\@brx\kern-0.5\wd\@brx\usebox{\@brx}}}
            \newcommand{\rrangle}[1][]{\savebox{\@brx}{\(\m@th{#1\rangle}\)}%
              \mathclose{\copy\@brx\kern-0.5\wd\@brx\usebox{\@brx}}}
            \makeatother        
            \newcounter{carrowover}

        \newcommand{\ul}[1]{{$\,$\underline{#1}$\,$}}

        \newcommand{\ignore}[1]{}
        \newcommand{\bibfnamefont}[1]{#1}
        \newcommand{\bibnamefont}[1]{#1}

        \newcommand{\tmax}{$t_{max}$}

\mathchardef\mhyphen="2D

\begin{document}
%\selectlanguage{English}

%\begin{CJK}{UTF8}{song}

%FRONTMATER     
    %TITLE    

        \title{Infinite temperature transport in the strong coupling regime of a nonintegrable quantum spin chain }
        %\title{ Anomalous transport in the strong coupling regime of a nonintegrable quantum spin chain }
        %\title{Bound states and anomalous transport of spinless ferimons with power-law interactions on one dimential lattices}

    %\author{Zhi-Hua Li$^{1}$ and An-Min Wang$^1$}
    %\address{$^{1}$Department of Modern Physics, University of Science and Technology of China, Hefei 230026, China}%\\
    \author{Zhi-Hua Li}
    \address{School of Science, Xi'an Technological University, Xi'an 710021, China}%\\

    \begin{abstract}
    We study spin transport of the XXZ model with next-nearest neighbor
    $\Delta_2$ terms. We compute numerically dependence of spin conductivity
    $\sigma(\omega)$ on the anisotropy $\Delta$ and the ratio $r=
    \Delta_2/\Delta$, in the large $\Delta$ regime. We find that, when $0<r<1$,
    the low-frequency conductivity assumes an anomalous form
    $\sigma(\omega)\approx a \omega^2 + b  \Delta^{-2} $. In particular, when
    $\Delta\to\infty$ the model becomes kinetically constrained and most states
    are localized.  We show, microscopically, existence of magnon bound states
    in the strong coupling regime, which behave as self-generated disorders for
    single magnons. Based on this quasiparticle picture, we obtain analytical
    scalings, which match well with the numerical results. 
    \end{abstract} 
    %%
        %\pacs{03.65.Ud, 75.10.Jm, 03.65.Fd}  

        % 75.10.Jm – Quantized spin models, including quantum spin frustration
        % 03.65.Ud – Quantum mechanics: Entanglement and quantum nonlocality (e.g. EPR paradox, Bell’s inequalities, GHZ states, etc.)
        % 02.20.Uw Quantum groups

        % 75.10.Pq: spin chain models, 
        % 03.65.Fd	Algebraic methods (see also 02.20.-a Group theory)
    \maketitle
    %\linenumbers
    \modulolinenumbers[5]

\paragraph*{Introduction.---}
%\section{introduction}

    Macroscopic hydrodynamics emerges from nonequilibrium many particle systems
    in the long-wave limit, which exhibits either normal diffusive transport or
    anomalous transport.  Understanding under what conditions one gets normal
    transport  is one of the main challenges of theoretical physics
    \cite{bertini2020finite}.  In recent years, this problem  has been
    undergoing intensive study, by investigating hight temperature transport in
    one-dimensional  quantum lattice models \cite{bertini2020finite}.  For
    integrable models, various anomalous transport is found
    \cite{castella1995integrability,zotos1996evidence,karrasch2013drude,karrasch2017hubbard,ljubotina2019kardar},
    and the microscopic mechanisms are nearly fully explained based on  
    existence of an extensive number of ballistic-moving quasiparticles
    %the picture of ballistic moving of quasiparticles
    \cite{prosen2013families,bertini2016transport,castroalvaredo2016emergent,ilievski2017microscopic,denardis2018hydrodynamic,bulchandani2021superdiffusion}. 
    %For the more practical nonintegrable systems, similar quasiparticle
    %pictures usually do not exist, and establishing a general framework to
    %understand their transport is formidable.
    For the more practical nonintegrable systems, similar quasiparticle
    pictures usually do not exist though.
    
    %It is desirable that for nonintegrable there could be
    %quasiparticle pictures. 
    %general lattice model, the Hamiltonian being $\hat H=\lambda \hat T +g\hat
    %V$. The weak coupling limit $g/\lambda \ll 1$ may be trivially understood by
    %perturbation theory. The on the other side of the spectrum $g/\lambda \gg 1$
    %is less trivial. In the latter limit, the particles can form bound states. 

    %Finding general mechanisms for general nonintegrable models is daunting. 
    %One practical way to proceed for nonintegrable models may be to restrict to
    %certain regimes of parameters in  certain types of models --- here we mean
    %the strong coupling regime of lattice models. In this regime, a salient
    %feature is that $n$ single particles can form bound states. This is because
    %that on a lattice single particles have a finite band width, then once a
    %cluster of particles gain a large potential-to-kinetic energy ratio (PKR),
    %they can not depart far away from each other afterwards. Note this only
    %relies on the principle of conservation of energy, so it holds generally for
    %integrable and nonintegrable models (an example for the latter is shown in
    %Fig.\ref{fig:intro} (a)).  The effective velocity generally decreases
    %exponentially with $n$, e.g. for the XXZ models it scales as $\sim
    %\Delta^{1-n}$ \cite{takahashi2005thermodynamics}. The bound states may jam
    %the motion of the entire system and lead to anomalous dynamics. 

\begin{figure}   
          \centering
          \subfigure{\scalebox{0.55}[0.55]{\includegraphics[width=80mm]{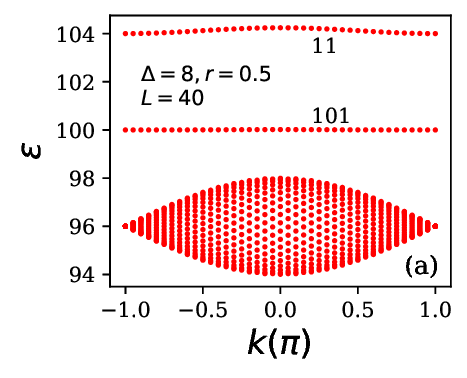}}}
          \raisebox{4mm}{\subfigure{\scalebox{1.05}[1.05]{\includegraphics[width=37mm]{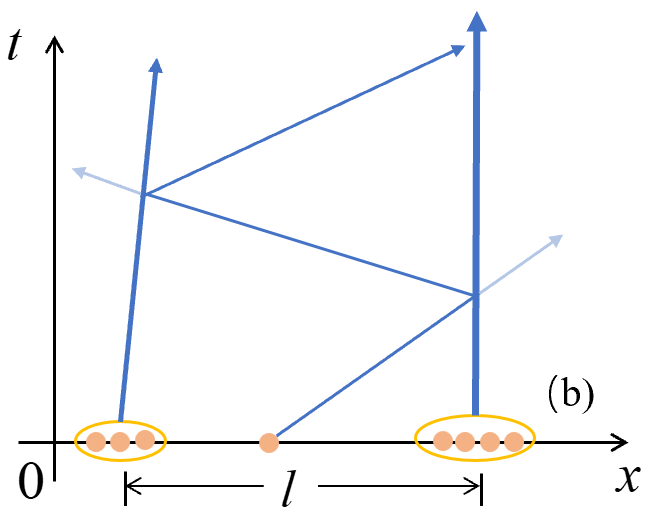}}}}
          %\raisebox{1mm}{\subfigure{\scalebox{1.0}[1.0]{\includegraphics[width=27mm]{ggg.png}}}}
          \caption{\label{fig:intro} 
          \textbf{A quasiparticle picture in the strong coupling regime.} 
          Panel (a): two-particle spectrum of model \eqref{eq:ham}. The two
          nearly flat bands on the top are bound states with configurations 11 and 101.
          Panel (b): For $0<r<1$, a fast-moving magnon is essentially confined
          between two slow-moving bound states. For finite $\Delta$, the
          magnon can transmit at certain rates; for
          infinite $\Delta$, the bound states are immobile, and the magnon is
          strictly confined.
        } 
    \end{figure}

    Given that establishing a theoretic framework for transport of general
    nonintegrable models is daunting, a practical way to proceed  may be to
    restrict to certain types of models and find their common mechanisms.  One
    such case, of particular interest, is the strong coupling regime of lattice
    models, where a cluster of $n$ particles with a large
    potential-to-kinetic energy ratio (PKR) will form bound states
    \cite{koster1954wave,kagan1984localization,mattis1986few,winkler2006repulsively,valiente2010three,fukuhara2013microscopic,kranzl2023observation,schmiedinghoff2022three,li2023slow}.  
    This is a general fact, valid for both  integrable or nonintegrable models [an
    example for the latter is shown in Fig.\ref{fig:intro} (a)]. 
    %In addition an $n$-particle bound state moves exponentially slow in $n$,
    %and, at large couplings, the only fast-moving objects are the singletons
    %(i.e. magnons for a spin model). The dynamics is then dominated by the
    %motion of singletons and their scatterings with the bound states, which
    %usually leads to nontrivial slow dynamics. This gives a quasiparticle
    %picture for the lattice models in the strong coupling regime. 
    In addition, an $n$-particle bound state moves exponentially slow in $n$.
    The dynamics is then dominated by the motion of singletons (i.e. magnons for
    a spin model) and their scatterings with the bound states. Therefore, not
    only there can be nontrivial slow dynamics \cite{kagan1984localization}, but also we have a quasiparticle
    picture for understanding it.

    %So there can be well-defined quasiparticles, and they exist generally for
    %integrable and nonintegrable lattice models [an example for the latter is
    %shown in Fig.\ref{fig:intro} (a)]. Furthermore, an $n$-particle bound state
    %moves exponentially slow in $n$, and, at large couplings, the only
    %fast-moving objects are the singletons (i.e.  magnons for a spin model). In
    %particular, for nonintegrable models the singletons may be almost
    %backscattered by the bound states [see Fig.\ref{fig:intro} (b)].  This can
    %lead to nontrivial slow dynamics. 
    
    % In this regime, a salient
    %feature is that $n$ single particles can form bound states. This is because
    %that on a lattice single particles have a finite band width, then once a
    %cluster of particles gain a large potential-to-kinetic energy ratio (PKR),
    %they can not depart far away from each other afterwards. Note this only
    %relies on the principle of conservation of energy, so it holds generally for
    %integrable and nonintegrable models (an example for the latter is shown in
    %Fig.\ref{fig:intro} (a)).  The effective velocity generally decreases
    %exponentially with $n$, e.g. for the XXZ models it scales as $\sim
    %\Delta^{1-n}$ \cite{takahashi2005thermodynamics}. The bound states may jam
    %the motion of the entire system and lead to anomalous dynamics. 

    %Because the particles in a lattice model having finite bind width, then
    %potential-to-kinetic energy ratio (PKR) can not depart far away because the
    %particles with large potential-to-kinetic energy ratio (PKR) can not depart
    %far away.  This fact relies on no assumption of integrability and maybe
    %holds
    
    %Indeed, in the past few years, various anomalous dynamics is found in the
    %strong coupling models, including the infinite coupling limit (which leads
    %to kinetically constrained models).
    Indeed, in the past few years, various anomalous dynamics is found in the
    strongly coupling models \cite{roeck2014asymptotic, schiulaz2014ideal,
    schiulaz2015dynamics, yao2016quasi,roeck2014asymptotic, barbiero2015out, yao2016quasi,
    abanin2017rigorous,mondaini2017many,
    bols2018asymptotic,michailidis2018slow, li2021hilbert,spielman2024quantum}  and the
    kinetically constrained models
    \cite{turner2018weak,horssen2015dynamics,moudgalya2020thermalization,brighi2023hilbert,yang2020hilbert,yang2022distinction,causer2024nonthermal}
    (the two are intimately connected, as the infinite coupling limit of the
    former usually leads to the latter \cite{dias2000exact,zadnik2021folded,zadnik2021foldeda}).  These studies
    are mostly focused on the problem of finding
    translation-invariant systems that violate the eigenstate thermalization
    hypotheses. To explain the phenomenon, new notions like Hilbert space
    fragmentation \cite{khemani2020localization,sala2020ergodicity} and quantum
    many body scars \cite{turner2018quantum,serbyn2021quantum} are  
    introduced. Transport in these types of models is also studied
    \cite{dias2000exact,sanchez2018anomalous, singh2021subdiffusion,nardis2022subdiffusive,chen2024superdiffusive,li2023slow}. 
    However, the above quasiparticle picture is rarely utilized
    \cite{nardis2022subdiffusive,li2023slow}.  To unveil the impacts of bound
    states on transport in nonintegrable models, in this work, we study spin
    transport of the XXZ model with next-nearest neighbor (NNN) terms. 
    %In this work, we study spin transport in one of the simplest nonintegrable
    %model, namely, the XXZ model with a next-nearest neighbor (NNN) $\Delta_2$
    %term. 
    We compute numerically the spin conductivity, and obtain analytical scalings
    based on the quasiparticle picture and perturbation theory.

    For infinite $\Delta$, we obtain the effective Hamiltonian  for the
    kinetically constrained model.  We find that when $r=0$ (absence of NNN
    terms) or $r=1$ (nearest and NNN terms are equal), single magnons can
    transmit all types of bound states. Otherwise, they  are completely blocked
    by most types of bound states [see Fig.\ref{fig:intro} (b)], with certain
    exceptions of unblocking states.  The former leads to a component of the
    conductivity which scales as $\sim \omega^2$, reminiscent that of Anderson
    localization \cite{mott1968conduction} or that deep in the many body
    localization \cite{gopalakrishnan2015low}. The unblocking states lead to a
    finite dc conductivity for finite systems,  which nevertheless vanishes in
    the thermodynamic limit.
   
    %We find that when $\Delta=\infty$  the system becomes dynamically
    %constrained. Whenever the range of interactions is larger than 1 (i.e.
    %$r\neq0$) and the strength of the nearest and next-nearest neighbor terms
    %are unequal (i.e. $r\neq1$), single magnons will be completely blocked by
    %most types of bound states, with certain exceptions of unblocking states.
    %The former leads to a component of the conductivity which scales as $\sim
    %\omega^2$, reminiscent that of Anderson localization
    %\cite{mott1968conduction} or that deep in the many body localization
    %\cite{gopalakrishnan2015low}. The unblocking states lead to a finite dc
    %conductivity for finite systems,  which nevertheless vanishes in the
    %thermodynamic limit.
    %The dynamics in the former subspace is essentially single-particle-like, 
    %The many body effects manifests mainly at finite $\Delta$, as a first order
    %correction in $1/\Delta$. 

    At finite $\Delta$, the dc conductivity receives a correction which  scales
    as $\sim \Delta^{-2}$. Microscopically, this is because  a magnon  can now
    transmit all types of bound states at finite rates.  The
    singleton-and-bound-state scatterings (SBSs) can be classified into two
    types: A magnon undergoes direction flip processes in the first type, while
    destruction/creation processes in the second type.  Among all SBSs, the
    transmission rate scales as $\sim \Delta^{-2}$ to the leading order, which
    gives a rough explanation of the correction in dc conductivity.

    %The transmission rate of a magnon through the former acting on $n$ sites
    %decays exponentially with $n$. Wheraz it has no
    %Last and most importantly, we point out a fundamental difference between the
    %disorder by random fields and the self-generated disorder by bound states:
    %The transmission rate of a singleton through a local field acting on $n$
    %consecutive sites decays exponentially with $n$, in contrast it has no
    %dependence on $n$ for a $n$-particle bound state (but it does has an
    %exponential dependence on the range ${R}$ of interactions).  Then our work
    %implies that there should be no transition in the type of transport at
    %\emph{finite} couplings for other clean nonintegrable lattice models with
    %$R\le 2$, but there could be a transition for even longer-ranged models. 

    %We point out a fundamental difference between the disorder by random fields
    %and the self-generated disorder by bound states.  Comparing transmission of
    %a singleton through the former acting on $n$ consecutive  sites and the
    %latter comprising $n$ particles, the rate of the former decays exponentially
    %with $n$, while the latter has no dependence on $n$.  (but it does have an
    %exponential dependence on the range $R$ of the potential energy terms).
    %This implies that in general clean models, for small $R$ there should be no
    %transition in the type of transport  at \emph{finite} couplings, but for
    %sufficiently large $R$ there may be a transition, i.e. there existing a
    %critical value $R_c$. The present results implies that $R_c>2$. 
  
    We point out a fundamental difference between the disorder by random fields
    and the self-generated disorder by bound states.  Comparing transmission of
    a magnon through a local field acting on $n$ consecutive  sites and  that
    through a bound state comprising $n$ particles at finite couplings, the rate of the former
    decays exponentially with $n$, while the latter has no dependence on $n$
    (but it does have an exponential dependence on the range  of the
    potential energy terms).  That is why there is a localization-delocalization
    transition at finite intensity of the random field, but the transition only
    occurs at infinite coupling for the clean model.

    %We also highlight two remarkable effects in the MBS.  For the type-I
    %scatterings, there is a destructive interference effect at $r=0.5$, making
    %the transmission rate vanish. For the type-II scatterings, there is a
    %tachyonic effect, that is a magnon can instantaneously transmit a bound
    %state with arbitrary widths.   
    
    %The dynamics is amenable to degenerate perturbation theories, specifically the
    %Schrieffer-Wolf transformation. 
    
    \comment{
         The question of how of closed quantum system thermalization is a fundamental
        question. Systems can be classified into ETH obeying and violating systems
        \cite{deutsch1991quantum,srednicki1994chaos}.  Notable examples are the MBL
        systems and integrable systems. Another important question is the emergent
        of hydrodynamics \cite{bertini2020finite}. Systems can be classified by
        normal diffusive transport and otherwise anomalous ones, labled by various
        dynamical exponents. A system that has difficulty in thermalization usually
        displays anomalous transport, a exact correspondence between the two is not
        fully established though. 
        
        An intriguing question is whether MBL-like phase and anomalous dynamics can be
        found in translation-invariant models. Many such kind of
        models are found, which typically fit into two categories: One
        category is the models with large potential-to-kinetic-energy ratios
        \cite{roeck2014asymptotic, grover2014quantum, schiulaz2014ideal,
        schiulaz2015dynamics,yao2016quasi,
        abanin2017rigorous,mondaini2017many,bols2018asymptotic,michailidis2018slow,spielman2022slow,yao2016quasi,roeck2014asymptotic,bols2018asymptotic}.
        Another category is systems with local dynamical constraints 
        \cite{turner2018quantum,horssen2015dynamics,brighi2023hilbert,yang2020hilbert}.
        %such as the PXP model,  quantum East model, Fredkin model and so on.  
        %such as the PXP models \cite{turner2018quantum}, quantum East model
        %\cite{horssen2015dynamics,brighi2023hilbert}, Fredkin model, and so on
        %\cite{yang2020hilbert}.  
        Although much progress has been made in this line of research, there are
        still several deficiencies.

        %%% kinetically constrained model; fragmentation; quantum manybody scars  
        
        First, the above two category of models are often studied separately.
        However, they can be intimately connected, because, if sending the potential
        to kinetic energy ratio to infinity, one usually obtains a kinetically 
        constrained model \cite{zadnik2021folded}. 
        %How does the dynamics change when the coupling changes from
        %infinity to finite values is less studied.  
        Second, in quantum lattice models, when the potential energy is larger than
        kinetic energy, there will be $n$-particle bound states solutions.  This is
        true both for integrable and nonintegrable models (an example for the latter
        is shown in Fig.\ref{fig:intro} (a)).  The effective velocity generally
        decreases exponentially with $n$, e.g. for the XXZ models it scales as $\sim
        \Delta^{1-n}$ \cite{takahashi2005thermodynamics}. Thus, at large couplings,
        the only fast-moving objects are the singletons (i.e. magnons for a spin
        model).  We advocate that, for clean \emph{nonintegrable} models, the bound
        states behave as self-generated disorders \cite{kagan1984localization,
        roeck2014asymptotic}, which is a third kind of disorder other than quenched
        disorder and random field. However, this quasiparticle picture has been
        seldomly used in existing works and deserves more attention. 
        %The ETH violation or anomalous transport are often explained by the
        %mechanisms like Hilbert space fragmentation or quantum many body scars.
        %Last, technically, the anomalous dynamics are often diagnosed by
        %entanglement, spectrum properties, and return probability. 
        Last, the optical conductivity, an important quantity in characterizing
        transport, has been calculated for disordered models.
        \cite{agarwal2015anomalous,gopalakrishnan2015low}. However, its behavior is
        unknown for most of the above two categories of models
        \cite{nardis2022subdiffusive}. 
       
        In this Letter, we study spin transport in the strong coupling regime of a
        nonintegrable spin chain, both in the finite coupling and infinite coupling limit
        (the latter leading to new kinetically constrained models), numerically
        calculate the optical conductivity, and explain the numerical
        results based on the just mentioned quasiparticle picture. We find that the
        infinite coupling limit of a nonintegrable model is not guaranteed to be
        thermal (conductor) or localized (insulator), which depends on whether the
        singletons can transmit most bound states or not. For the latter case the
        system is in a localized state with the conductivity scaling as $\sim
        \omega^2$, reminiscent of Mott's law of Anderson localization
        \cite{mott1968conduction}; when the coupling coefficient $\Delta$ is finite, the dc
        conductivity becomes finite and scales as $/\Delta^{-2}$, because of a finite
        transmission rate with the same scaling.
        
        One very remarkable fact is that the just-stated transmission of a singleton through an
        $n$-particle bound state depends only on the coupling strength, but
        \emph{not} on $n$. 
        %This is in stark contrast with the disorder by random
        %field, where transmission rate through a potential barrier of length $n$
        %scales as $e^{-n}$. 
        This implies that, for generic nonintegrable
        models, whenever the density of magnons is finite, there will be a finite dc
        conductivity at finite coupling (i.e. excluding true localization-delocalization
        transition at finite coupling). 
    }

    %systems, which is coined generalized hydrodynamics (GHD)\cite{bertini2016transport,castroalvaredo2016emergent}. 

    %The rest of the paper is organized as follows: Sec. \ref{sec:model}
    %introduces the model Hamiltonian and observables.  Sec. \ref{sec:results}
    %presents numerical results demonstrating the slow transport and relaxation
    %properties. Sec. \ref{sec:bound:state} delivers a systematic study of the
    %bound states of the model, including properties of their spectra, group
    %velocities and scatterings, based on which the transport and relaxation
    %processes are interpreted.  Finally, conclusions are drawn in
    %Sec. \ref{sec:conclusion}.  

\paragraph*{Model and observable.---}        \label{sec:model}
    The model considered is the XXZ spin chain with NNN interaction terms 
    \begin{equation} 
        %H = \sum\limits_i {J(S_i^xS_{i + 1}^x + S_i^yS_{i + 1}^y) + \Delta S_i^zS_{i + 1}^z + \Delta_2 S_i^zS_{i + 2}^z}. 
        H = \sum_{i=1}^L {J(S_i^xS_{i + 1}^x + S_i^yS_{i + 1}^y) + \Delta
        (S_i^zS_{i + 1}^z + r S_i^zS_{i + 2}^z}), 
        \label{eq:ham}
    \end{equation} 
    It is nonintegrable when $\Delta_2\equiv r\Delta\neq 0$, and equivalent to a
    system of spinless fermions with NNN interactions, known
    as the t-V-W model \cite{giamarchi2004quantum}. 
    Then up and down spins are identified with particles and empty sites, 
    and the notation $1=\uparrow$ and $0=\downarrow$ is adopted.  
    %As stated in the above, when $J<\Delta$, bound states exist. 
    %We restrict our discussion to $\Delta$ being at least several multiples of
    %$J$, where the transport is dominated by single magnons and their
    %scatterings with the bound states, and also restrict $0\le r\le 1$, namely,
    %the interactions being convex.
    We assume $0\le r\le 1$, i.e. convex interactions. And we restrict our  
    investigation on the high-temperature transport to $\Delta$ being at least
    several multiples of $J$, where bound states play important roles. 
    %$J$
    For $\Delta_2=0$, viz. the XXZ model, each bound state amounts to $n$
    consecutive flipped spins in a background of down spins, known as  an
    $n$-string. For $\Delta_2\neq 0$, new types of bound states emerge,  the
    spin configurations being like 101, 1101, 1011, 11101$ \cdots$
    \footnote{At infinite $\Delta$, these classical spin configurations are
    exact eigenstates. So they are used to denote each bound states. At finite
    $\Delta$, they are approximately eigenstates, with the amplitudes of
    tunneling to other configurations at the order of $O(1/\Delta)$. Therefore,
    for the latter case (especially at large $\Delta$), it is still legible to
    use the bit-strings to denote specific types of bound states.}.  
    For convenience, we call the \emph{classical} spin configuration 111 as a
    $(3)$-block, 11101 as a $(3+1)$-block, and so forth. 
   
    %\footnote{The $(n+1)$-string and $(1+n)$-string are resonant in energy, and
    %the proper eigenstates are superposition of the two with definite parities,
    %like $(|1101\rangle \pm  |1011\rangle)/\sqrt{2}$}. 
    
    %The  equilibrium states of this model at zero temperature  have been
    %extensively  studied (see \cite{mishra2011phase} and references therein).
    %Here we study its transport at infinite temperature. 
    The infinite temperature transport of the model can be characterized by the
    spin conductivity, the real part being  
    \begin{equation}
        T\sigma (\omega ) = Re\int_0^\infty  { \frac{1}{L} \langle
        \hat J(t) \hat J(0)\rangle {e^{i\omega t}}d} t,
        \label{eq:sigma}
    \end{equation}
    where  $\hat J = \sum_i {J(S_i^xS_{i + 1}^y - S_i^yS_{i + 1}^x)}$ is the
    spin current operator, $\hat J(t)=e^{-i H t/\hbar} \hat J e^{i H t/\hbar}
    $ and $\langle \cdot \rangle=\frac{1}{2^L}\text{Tr}(\cdot)$ denotes average
    in the grand canonical ensemble. Although $\sigma\to 0$ when the temperature
    $T\to\infty$, the product $T\sigma$ remains finite.  Hereafter, $\sigma$ is
    interpreted as $T\sigma$,  and the units $J=\hbar=1$ are
    used. 
    
    The dynamical quantum typicality (DQT) method
    \cite{steinigeweg2014spin,steinigeweg2015spin}  is employed to numerically
    evaluate the correlator $C(t)=\langle \hat J(t)
    \hat J(0)\rangle/L$.
    Then a cut-off by a $t_{max}$ in the temporal integral of
    Eq.\eqref{eq:sigma} has to be made, which is set to range from 14 to 20 (see  
    %\cite{supplementary2024}
    Supp. Mat. Sec. \ref{sec:supp:numerics:tmax} for details). 
    %The lattice momentum $k$ and the magnetization $m$ are conserved, and
    %invariant under the action of the current operator.  By exploiting these
    %symmetries, we evaluate $C(t)$ for up to $L=34$ spins in the $k=0$ sector
    %with periodical boundary conditions (PBC).  
    By exploiting conservations of lattice momentum $k$  and magnetization $m$,
    we evaluate $C(t)$ for $L$ ranging from 24 to 34 in the $k=0$ sector with
    periodical boundary conditions (PBC) (see Supp. Mat. 
    Sec. \ref{sec:supp:numerics:symm} and  \cite{steinigeweg2015spin}).

\paragraph*{The $\Delta=\infty$ limit}  \label{sec:results}
    
    %%%
        %\begin{figure}   % phase diagram 
        %      \centering
        %      \scalebox{0.55}[0.55]{\includegraphics{bowling_inf_v.eps}}
        %      \caption{\label{fig:sz:y:t} 
        %      Time evolution of the spin density $\mu^{-1} m(i,t)$ at $V=4$
        %      (left panel) and $V=16$ (right panel) on a chain with $L=256$ sites.
        %      The origins of the $i$-axes are shifted to $L/2$.  
        %      }
        %      \end{figure}

    \begin{table}[h]  %table 里面也可以嵌套tabular,只有tabular是不能加标题的
        \centering  %表格居中
        \caption{\textbf{Local dynamical constraints on hopping of spins}. 
        Hopping of the spin at the third site to the fourth site (and back)
        depends on four neighboring spins (underlined sites) and on $r$.  The
        values of $r$ are grouped together when the allowed hoppings are the
        same, leading to four distinct groups.
            }  
    \begin{small}
        %\begin{tabular}{p{3cm}lll}
        %\begin{tabular}{p{2.0cm}p{2.0cm}p{2.0cm}p{2.0cm}p{2.0cm}}
        \begin{tabular}{p{1.6cm}p{0.6cm}cp {1.0cm}p{0.5cm}}
        %\centering
        \toprule  % 顶部线
          $\quad r$   & 0 &  \small{$(0,0.5)\cup(0.5,1)\quad$} &  $0.5$  & $1$ \\
        \hline
        \ul{00}10\ul{00} & \checkmark & \checkmark & \checkmark & \checkmark \\
        \hline                                     
        \ul{00}10\ul{01} & \checkmark &            &            &            \\
        \hline                                     
        \ul{00}10\ul{10} &            &            &            & \checkmark \\
        \hline                                     
        \ul{00}10\ul{11} &            &            &            &            \\
        \hline                                     
        \ul{01}10\ul{00} &            &            &            & \checkmark \\
        \hline                                     
        \ul{01}10\ul{01} &            &            & \checkmark &            \\
        \hline                                     
        \ul{01}10\ul{10} & \checkmark & \checkmark & \checkmark & \checkmark \\
        \hline                                     
        \ul{01}10\ul{11} & \checkmark &            &            &            \\
                                                   
        \hline                                     
        \ul{10}10\ul{00} & \checkmark &            &            &            \\
        \hline                                     
        \ul{10}10\ul{01} & \checkmark & \checkmark & \checkmark & \checkmark \\
        \hline                                     
        \ul{10}10\ul{10} &            &            & \checkmark &            \\
        \hline                                     
        \ul{10}10\ul{11} &            &            &            & \checkmark \\
        \hline                                     
        \ul{11}10\ul{00} &            &            &            &            \\
        \hline                                     
        \ul{11}10\ul{01} &            &            &            & \checkmark \\
        \hline                                     
        \ul{11}10\ul{10} & \checkmark &            &            &            \\
        \hline                                     
        \ul{11}10\ul{11} & \checkmark & \checkmark & \checkmark & \checkmark \\
        
        \bottomrule  % 顶部线
        
        %\specialrule{0em}{0pt}{0pt}   % 命令第一个大括号控制表格线的粗细，若为0，则表格线透明，第二个大括号是表格线与上方内容的距离，第三个大括号是表格线与下方内容的距离，通过改变后两个大括号中的值来控制行高！
        %\end{tabular} 
    \end{tabular}
    \label{tab:allowed:hopping}
    \end{small}
    \end{table}
    
    \begin{figure}   % phase diagram 
          \centering
          \scalebox{0.55}[0.55]{\includegraphics{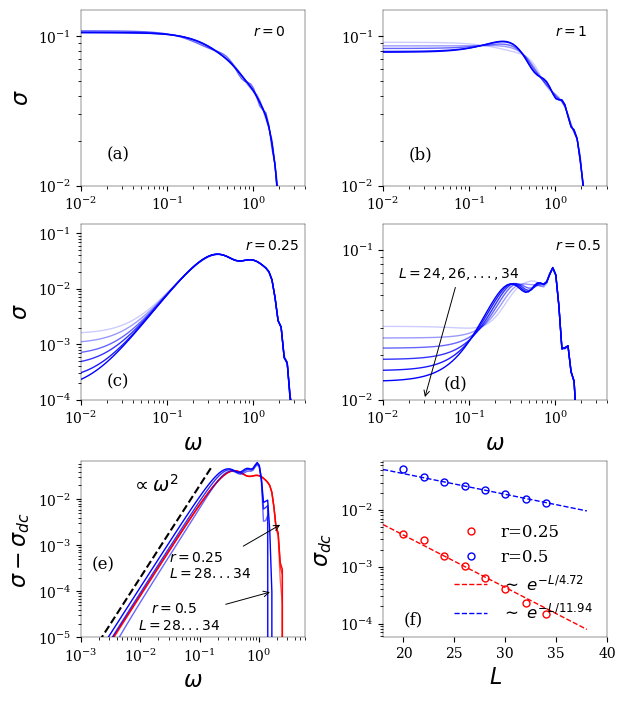}}
          \caption{\label{fig:sigma:inf:V} 
          %\textbf{Conductivity in the $\Delta\to\infty$ limit.}
          Panels (a)-(d): Spin conductivity for the effective
          Hamiltonian $\tilde H_r^{(0)}$ in a  log-log scale for $r=0$, 1, 0.25, and
          0.5, respectively (high-frequency weights vanish for
          $\omega\gtrsim 2$  in all cases).  
          %Unblocking (blocking) of the magnon is related
          %to the relatively higher (lower) low-frequency conductivity  in panel
          %(b) [panel (c)]. 
          Panel (e): For $r=0.25$ and 0.5, after extracting the dc conductivity,
          the residual scales as $\sim \omega^2$. Panel (f): The
          finite-size dc conductivity $\sigma_{dc}(L)$ for $r=0.25$ and 0.5 can
          be fitted by the exponential functions $\sim e^{-L/\lambda}$, with
          $\lambda=4.72$ and $11.94$, respectively. 
          %obtained through DQT for lattice sizes $L=24, 26, \dots,
          %34$ in the $k=0$ sector. 
          }
          \end{figure}

    In this limit, the quantity \begin{equation} \label{eq:conserve:law}
        B=N_{11} + r N_{1*1} \end{equation} is conserved, where
        $N_{11}=\sum_{i=1}^{L}\langle \hat{n}_i \hat{n}_{i+1} \rangle$ and
        $N_{1*1}=\sum_{i=1}^{L}\langle \hat{n}_i \hat{n}_{i+2}\rangle$ counts number of
    nearest and next-nearest neighbored pairs of up spins, since changing $B$ costs
    infinite energy. 
    %This local conservation law constraints hopping of
    %spins between two neighboring sites $i$ and $i+1$, which depends on four
    %further neighboring sites, two on each sides: $s_{i-2} s_{i-1}
    %\carrowover{1_i0}_{i+1} s_{i+2} s_{i+3} $, and also depends on $r$.  It is
    %easy to see that $r\in[0,1]$  should be divided into four classes: 0, 1, 0.5
    %and $(0, 0.5)\cup (0.5,1)$, as summarized in Table \ref{tab:allowed:hopping}
    %(since for $r\in (0, 0.5)\cup (0.5,1)$ the allowed hoppings are the same,
    %subsequently we use $r=0.25$ as a representative for this interval).  
    %%
    %This local conservation law constraints hopping of spins between two
    %neighboring sites, which depends on four further neighboring sites and 
    %on $r$ as well, as summarized in Tab. \ref{tab:allowed:hopping}. The values of
    %$r$ are classified into four classes. In each of them, the allowed hoppings
    %are the same. 
    This local conservation law constraints hopping of spins at two
    neighboring sites $i$ and $i+1$, which depends on four further neighboring sites and on
    $r$ as well, as
    summarized in Tab. \ref{tab:allowed:hopping}. 
    Alternatively, the constraints are compactly represented by
    projectors acting on the four neighboring sites, for each group of $r$,  
    \begin{equation}
        \label{eq:projector}
        P_{r,i} = 
        \left\{ 
            \begin{array}{l}
                \frac{{\sigma _{i - 1}^z\sigma _{i + 2}^z + 1}}{2},\quad  r = 0\\
                \frac{{\sigma _{i - 2}^z\sigma _{i + 3}^z + 1}}{2},\quad  r = 1\\
                \frac{{(\sigma _{i - 1}^z\sigma _{i + 2}^z + 1)}}{2}\frac{{(\sigma _{i - 2}^z\sigma _{i + 3}^z + 1)}}{2},\quad r \in (0,0.5)\cup(0.5,1) \\
                %\frac{3}{8}\sigma _{i - 2}^z\sigma _{i - 1}^z\sigma _{i + 2}^z\sigma _{i + 3}^z - \frac{1}{8}\sigma _{i - 2}^z\sigma _{i - 1}^z \\ 
                %       - \frac{1}{8}\sigma _{i + 2}^z\sigma _{i + 3}^z  + \frac{1}{8}\sigma _{i - 2}^z\sigma _{i + 2}^z  + \frac{1}{8}\sigma _{i - 2}^z\sigma _{i + 3}^z  \\
                %        + \frac{1}{8}\sigma _{i - 1}^z\sigma _{i + 2}^z + \frac{1}{8}\sigma _{i - 1}^z\sigma _{i + 3}^z + \frac{3}{8},\quad \quad r = 0.5
                \frac{1}{2}(\frac{{\sigma _{i - 2}^z\sigma _{i - 1}^z - 1}}{2}\frac{{\sigma _{i + 2}^z\sigma _{i + 3}^z - 1}}{2} + \frac{{\sigma _{i - 2}^z\sigma _{i + 2}^z + 1}}{2}\frac{{\sigma _{i - 1}^z\sigma _{i + 3}^z + 1}}{2} \\
                + \frac{{\sigma _{i - 1}^z\sigma _{i + 2}^z +
                1}}{2}\frac{{\sigma _{i - 2}^z\sigma _{i + 3}^z + 1}}{2}), \quad
                r=0.5,
            \end{array} 
        \right.
    \end{equation}
    Then one readily writes down the effective Hamiltonian in the zeroth order 
        \begin{equation}
            \label{eq:ham:eff}
            {\tilde H_r^{(0)}} = J\sum\limits_i {{P_{r,i}}(S_i^xS_{i + 1}^x + S_i^yS_{i + 1}^y)} 
        \end{equation}
    Following the terminology used in
    Refs.\cite{zadnik2021folded,zadnik2021foldeda}, we call 
    it the folded t-V-W  model. 
    Since for $r \in (0,0.5)\cup(0.5,1)$ the
    effective Hamiltonians are identical, hereafter we use $r=0.25$ to
    represent this group. 
    
    We calculate the spin conductivity at $\Delta=\infty$
    directly via the effective Hamiltonian \footnote{For the effective
    Hamiltonian, the spin current operator is rewritten as 
    $\hat J = J\sum_i {P_{r,i}(S_i^xS_{i + 1}^y - S_i^yS_{i + 1}^x)}$.}. 
    %To calculate its
    %conductivity, the spin current operator is rewritten as 
    %$\hat J = J\sum_i {P_{r,i}(S_i^xS_{i + 1}^y - S_i^yS_{i + 1}^x)}$.
    The result for each $r$ is summarized in Fig.\ref{fig:sigma:inf:V}
    (a)-(d). For $r=0$ and 1 [panels (a) and (b)], the low-frequency
    conductivity barely depends on system sizes, which suggests the dc
    conductivity being finite in the thermodynamic limit. For $r=0.25$ and 0.5 [panels(c) and
    (d)], although the finite-size dc conductivity is finite, it is much lower
    and reduces as the system size increases. 
    %The different behavior is apparently
    %related to the scattering behavior of magnons with the bound states: For the
    %former two $r$ values, the magnons can transmit \emph{all} bound states [an
    %example is illustrated in panel (e) for $r=1$]. For the later two cases, the
    %magnons are completely reflected by \emph{most} bound states [an example is
    %illustrated for $r=0.25$ in panel (f)]. 
    
    The finite dc conductivity at $r=0$ and $1$ is related to the fact that a
    magnon can transmit all bound states. For example, a magnon (the underlined
    bit) can 
    transmit a $(5)$-block by the following process at $r=0$
    \begin{equation}
        %{\fontspec[FakeBold=1.2]{SomeFont} This is bold text.}
    \begin{split}
        \quad\,         &{ \underline{1}00111110 }    \quad\quad         \\
         \rightarrow     &\red{0\underline{1}0111110}      \quad\quad  \,\,\,\,\,\,r\Delta               \\
         \rightarrow     &011\underline{0}11110      \quad\quad  -r\Delta              \\
         \rightarrow     &0111\underline{0}1110      \quad\quad  \,\,\,\,\,\,0                     \\
         \rightarrow     &01111\underline{0}110      \quad\quad  \,\,\,\,\,\,0                     \\
         \rightarrow     &\red{0111110\underline{1}0}      \quad\quad  \,\,\,\,\,\,r\Delta               \\
         \,\,\rightarrow &01111100\underline{1}  \quad\quad  -r\Delta, 
         \label{eq:transmit:r:0}
    \end{split}
    \end{equation}
     because costs of energy in each step (indicated on the right for
    general $r$ and $\Delta$) become zero  when $r=0$. 
    At $r=1$, complete transmission is also possible, nevertheless via a
    slightly different process (differences are marked in red)
    \begin{equation}
    \begin{split}
         \quad\,       &\underline{1}00111110                                              \\
         \rightarrow   &\red{\underline{1}01011110}    \quad\quad  (r-1)\Delta               \\
         \rightarrow   &011\underline{0}11110     \quad\quad  (1-r)\Delta              \\
         \rightarrow   &0111\underline{0}1110      \quad\quad  0                       \\
         \rightarrow   &01111\underline{0}110      \quad\quad  0                       \\
         \rightarrow   &\red{01111010\underline{1}}      \quad\quad  (r-1)\Delta             \\
         \,\rightarrow &01111100\underline{1}     \quad\quad  (1-r)\Delta.            
         \label{eq:transmit:r:1}
    \end{split}    \end{equation}
    %at the same time the bound state is shifted to the left by two sites. 
    Note that in both processes, the magnon flips direction when it passes an
    domain wall \cite{vlijm2015quasi}. This leads to diffusive spin transport instead of
    ballistic transport
    \cite{gopalakrishnan2019kinetic,bulchandani2021superdiffusion}. 
    
    In contrast, at $r=0.25$ or 0.5, single magnons are completely blocked by
    most types of bound states, with few exceptions, which can be checked
    through the rules in Tab.\ref{tab:allowed:hopping}.  In the rest of the
    paper we focus only on this more interesting \emph{nearly} magnon-blocking regime
    of $0<r<1$, where the Hilbert space can be divided into two subspaces
    $\mathcal{H}=\mathcal{H}_{block}\oplus \mathcal{H}_{unblock}$. States in the
    former contain at least one magnon-blocking bound state, while in the latter
    there are purely unblocking states. 
    %In the rest of the paper we focus only on the more interesting cases of
    %$r=0.25$ and 0.5 (i.e. $0<r<1$).  Note that there are still certain types of
    %bound states that do not block the magnons, and the Hilbert space can be
    %divided into two subspaces $\mathcal{H}=\mathcal{H}_{block}\oplus
    %\mathcal{H}_{unblock}$. 
    Then an elaboration on the conductivity is made in
    panel (e), which shows that the finite-size conductivity behaves as
    $\sigma(L)=\sigma_{dc}(L) + a\omega^2$ at low frequency. The second
    component is universal,
     barely depending on $r$ or $L$, while the dc part has a clear 
     dependence on them [see panel (f)]. Below we demonstrate that these two
    components are attributed to the blocking and unblocking subspace, respectively,
    and make an analytical account of their scalings.

    For the states in the blocking subspace,  we have the picture
    that every magnon is trapped between two adjacent  bound states, with
    certain distance $l$ [see Fig.\ref{fig:intro} (b)].
    This is locally identical to a single particle moving on a finite chain of
    length $l$ with open boundary conditions (OBC). Conductivity for this
    single particle has a peak located at  $\omega^* \sim 1/l$, and 
    the peak value scales as $\sim1/l^2$, i.e. $\sigma(\omega^*)\sim
    \omega^{*2}$ (see Supp. Mat. Sec. \ref{sec:supp:single:particle:OBC} and 
    \cite{rigol2008drude}). 
    %Then for different magnons, which are confined various different length
    %$l's$, leading to the above scaling $\sigma(\omega) \sim \omega^2$. 
    Then for different magnons, which are trapped in all different length
    $l's$, we get the above scaling $\sigma(\omega) \sim \omega^2$.

    %Then we discuss how the unblocking subspace contributes to $\sigma_{dc}(L)$.
    %For $r=0.25$, $\mathcal{H}_{unblock}$ is spanned by all the computational
    %basis states that do not contain two consecutive bits of 1's.
    %$\sigma_{dc}(L)$ is approximately proportional to the ratio of the
    %dimensions of this subspace and the entire Hilbert space $\sigma_{dc}(L)
    %\sim dim{(\mathcal{H}_{unblock})}/2^L$. The former dimension is equal to the
    %Fibonacci number $\text{Fib}(L+2)$. For large $L$,  $\text{Fib}(L+2) \approx
    %\phi^{L+2}/\sqrt{5}$, where $\phi=(1+\sqrt{5})/2$ is the golden ratio.  Then
    %we get $\sigma_{dc}(L) \sim e^{-L/\lambda}$, with
    %$\lambda=1/ln(2/\phi)\approx 4.72$. The numerical data can be fitted well
    %with this function, as shown in panel (h). Similar treatment can be made for
    %$r=0.5$ as well. The unblocking subspace is spanned by all computational
    %basis states which do not contain three consecutive bits of 1's. The
    %dimension of this subspace equals to the Tribonacci number $\text{Trib}(L+3)$,
    %which scales as $\sim \rho^{L+3}$ with $\rho\approx 1.8393$ for large $L$ 
    %\cite{spickerman1982binet}. Then we obtain $\sigma_{dc} \sim e^{-L/\lambda}$
    %with $\lambda=1/ln(2/\rho)\approx 11.94$, which, again, matches well with the
    %numerical data. So, we conclude that, in the range of $0<r<1$,
    %$\mathcal{H}_{unblock}$ is only a vanishingly small fraction of the entire
    %Hilbert space, and that the system is localized in the thermodynamic limit. 

    Then we discuss how the unblocking subspace contributes to $\sigma_{dc}(L)$. 
    $\sigma_{dc}(L)$ is approximately proportional to the
    ratio of the dimensions of this subspace and the entire Hilbert space
    $\sigma_{dc}(L) \sim dim{(\mathcal{H}_{unblock})}/2^L$.   
    For $r=0.25$ (resp., 0.5), $\mathcal{H}_{unblock}$ is spanned by all the
    computational basis states that do not contain two (resp., three)
    consecutive bits of 1's,  whose dimension is equal to
    the Fibonacci number $\text{Fib}(L+2)$ (resp., the Tribonacci number
    $\text{Trib}(L+3)$).  At large $L$,
    $\text{Fib}(L+2) \approx \phi^{L+2}/\sqrt{5}$, where $\phi=(1+\sqrt{5})/2$
    is the golden ratio, and $\text{Trib}(L+3)$, which scales as $\sim
    \rho^{L+3}$ with $\rho\approx 1.8393$ \cite{spickerman1982binet}.  Then we
    get $\sigma_{dc}(L) \sim e^{-L/\lambda}$, with $\lambda=1/ln(2/\phi)\approx 
    4.72$ for $r=0.25$ and $\lambda=1/ln(2/\rho)\approx
    11.94$ for $r=0.5$. Fig.\ref{fig:sigma:inf:V} (h) shows that the numerical
    data can be well fitted with these analytical scalings.  So, we conclude that, in the
    range of $0<r<1$, $\mathcal{H}_{unblock}$ is only a vanishingly small
    fraction of the entire Hilbert space, and that the system is localized in
    the thermodynamic limit.  For convenience of later discussions, we also
    denote the above dc conductivity at $\Delta=\infty$ by $\sigma^{(0)}_{dc}$
    (meaning zeroth order in $1/\Delta$).

\paragraph*{Finite $\Delta$.---} \label{sec:bound:state}

    %%%
    \begin{figure}  
          \centering
          \scalebox{0.55}[0.55]{\includegraphics{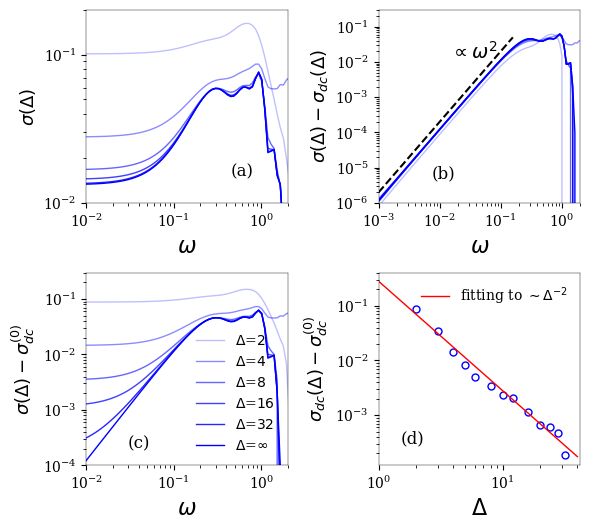}}
          \caption{\label{fig:sigma:finite:V} 
          Panel (a): Low-frequency conductivity for different anisotropy
          $\Delta$ [see the legend in panel (c)] at $r=0.5$, obtained for the
          maximal size $L=34$.
          Panels (b) and (c): Residual conductivity after extracting the dc part
          $\sigma_{dc}(\Delta)$ and a constant value
          $\sigma_{dc}^{(0)}=0.015$, respectively.  
          %The low-frequency regime of the former scales as as $\sim\omega^2$.  
          Panel (d): The dc part of the latter vs. $\Delta$, which can be fitted
          by a $\sim\Delta^{-2}$ scaling. 
          }
          \end{figure}
    
     %%%

    For finite $\Delta$, the quantity $B$ is not strictly conserved.  Changing
    $B$ can lead to up to four peaks in the conductivity at intermediate and
    high frequencies (see Supp. Fig. \ref{fig:supp:peak:pos}).  Below, we
    consider the hydrodynamic regime at low frequencies, and focus on $r=0.5$ in
    particular. The results should be qualitatively the same for other $r$
    values in the range of $0<r<1$, which we demonstrate in Supp. Mat.  Sec.
    \ref{sec:supp:sigma:low:freq}. 
    %Besides, we restrict the discussions only to the possibly magnon-blocking
    %regime $0<r<1$, focusing on $r=0.5$ in particular. 

    Figure\ref{fig:sigma:finite:V} (a) shows dependence of the conductivity on
    each $\Delta$ (including $\Delta=\infty$ for comparison). The
    finite-$\Delta$ effect seems to be weak, when $\Delta\gtrsim 8$.  If dc
    conductivity $\sigma_{dc}$ is subtracted, then the residue scales as
    $\sim\omega^2$ and the effect of $\Delta$ nearly completely disappear [see
    panel (b)].  This indicates that the only finite-$\Delta$ effect is this
    contribution to $\sigma_{dc}$.  While if, only
    $\sigma^{(0)}_{dc}\equiv\sigma_{dc}(\infty)$ is subtracted, the difference
    $\sigma(\Delta)- \sigma^{(0)}_{dc}$ has a salient dependence on $\Delta$
    [see panel (c)] and its dc part can be approximately fitted by the $\sim
    \Delta^{-2}$ scaling [see panel (d)].  
    %So, for finite systems we have $\sigma_{dc}(\Delta) \approx
    %\sigma^{(0)}_{dc} + a\Delta^{-2}$, the second term being recognized as the
    %first order contribution $\sigma^{(1)}_{dc}$ in $1/\Delta$.  In the
    %thermodynamic limit, the zeroth order term scales to zero, while the first
    %order term should survive. 
    So, for finite systems we have $\sigma_{dc}(\Delta) \approx
    \sigma^{(0)}_{dc} + b\Delta^{-2}$. In the thermodynamic limit, the first
    term scales to zero, while the second term should survive. 
    
    The second term is recognized the first order correction 
    $\sigma^{(1)}_{dc}$ to the dc conductivity. Microscopically, this is due to single
    magnons can now transmit all types of bound states at finite rates. The
    $\sim \Delta^{-2}$ scaling corresponds to an
    effective hopping amplitude $\sim 1/\Delta$ in the first order correction to
    the effective Hamiltonian.
    %The fact that  $\sigma^{(1)}_{dc}\sim \Delta^{-2}$ is roughly due to an
    %effective hopping amplitude $\sim 1/\Delta$ in the first order correction to
    %the effective Hamiltonian. 
    Instead of obtaining an explicit form for the effective Hamiltonian, here we
    analyze the microscopic SBSs, which contribute to $\sigma_{dc}$.  We identify
    two basic types (I and II) of SBSs, which are of rather different characters.
    The type-I SBSs are featured by direction flip of single magnons during the
    transmission. There are in turn two different routes as already exemplified
    by Eqs.  \eqref{eq:transmit:r:0} and \eqref{eq:transmit:r:1}.  The
    transmission rates are, respectively, $(r\Delta)^{-4}$ and
    $[(1-r)\Delta]^{-4}$, valid for $0<r<1$ and large $\Delta$. In the type-II
    scatterings, a magnon  is destroyed on one side of a bound state and created
    on the other side, which amounts to transmission (for identical particles).
    The rates depend on specific types of the bound states. For $(n+1)$-blocks, 
    the process (say $n=5$)
    \begin{equation}
    \begin{split}
            &\underline{1}0011111010    \\
        \to &0\underline{1}011111010   \quad\quad  \,\,\,\,r\Delta     \\
        \to &0101111100\underline{1}    \quad\quad -r\Delta   
         \label{eq:transmit:type:B:1}
    \end{split}
    \end{equation}
    has a rate $\sim (r\Delta)^{-2}$. 
    The rates for scatterings with other blocks may be analyzed likewise. 
    
    From the above analysis, we recognize the singleton-and-$(n+1)$-block scatterings 
     as the first order processes (other processes like
    Eqs.\eqref{eq:transmit:r:0} and \eqref{eq:transmit:r:1} are all of higher
    orders).  Each of these processes leads to a matrix element in the current correlation
    function scaling as $~\Delta^{-2}$ at $\omega\approx 0$. Supposing finite
    density of the $(n+1)$ type bound states in the system, this gives a rough
    explanation of the behavior of $\sigma^{(1)}_{dc}$.
    Two final remarks about the SBS follow. First, $r=0.5$ is a special symmetric
    point, for which the processes of Eqs.\eqref{eq:transmit:r:0} and
    \eqref{eq:transmit:r:1} are different by a phase of $\pi$. Then there should
    be a remarkable destructive interference effect making the type-I
    transmission absent at
    $r=0.5$. Second, in all the processes, the transmission rate has no
    dependence on $n$ for an $n$-particle bound state, which is obviously seen
    from the above analysis. 

\paragraph*{Summary.---}  \label{sec:conclusion}
%\paragraph*{Discussion.---}  \label{sec:conclusion}
    
    In summary, we studied infinite temperature transport of the XXZ model with
    NNN  terms.   We found that the strong
    coupling regime exhibits rich transport behaviors due to existence of bound
    states. For infinite couplings, the system can be either
    diffusive (at $r=0$ or 1) or localized ($0<r<1$). For the latter case, there
    is also coexistence of conducting states and insulating states, albeit the
    fraction of the conducting states is vanishingly small. For finite couplings,
    the conductivity receives corrections in low frequencies, and displays peaks
    at intermediate and high frequencies.  
    Given that bound states exist in general {nonintegrable} lattice models
    in the large PKR regime, we argue that similar behaviors may prevail. 
    The low-frequency conductivity may be usually in the form
    $\sigma(\omega)=\sigma^{(0)}_{dc} + a\omega^\alpha + b\Delta^{-\beta}$,
    where  $\Delta$ denotes the overall coupling of the potential energy terms,
    and $\sigma^{(0)}_{dc}$ denote the  dc conductivity  in the limit
    $1/\Delta\to 0$. 
    We expect the exponent $\alpha=2$ being universal and the exponent $\beta$
    to be model specific.

\section*{acknowledgement}    

     The author thanks Chun Chen for discussions. Our implementation of
     the DQT algorithm is based on the Quspin package
     \cite{weinberg2017quspin,weinberg2019quspin}.

%\section*{acknowledgement}    
%
%     The author thanks prof. *** for help with computation
%     resources. This work was supported by National Natural Science Foundation
%     of China under Grant No. 11375168 and by Research Starting Fund of Xi'an
%     Technological University.
%     %No. ????????.    
%     %This work is supported by National Natural Science Foundation of China
%     %under Grant 
%     %No. 11375168 and No.  
%     %thank pfeifer \\
    
%BIBLIOGRAPHY

%\appendix*

\section*{References}
   % when submit paper, using  .bbl file 
    %\bibliography{mps-wigner-crystal.bib, mps-wigner-crystal-2.bib}
    %\bibliography{C:/Users/zhihua/Dropbox/my-bib-for-all-paper.bib}   % use ~/Dropbox/... cause error   
    %\bibitem{S_RefA} A. Someone, C. Someone, D. Someone, Phys. Rev. Lett. {\bf 11}, 1111 (1911).
    
    %\begin{thebibliography}{./q-lmg-bibliography}
    %\end{thebibliography}

%\appendix

%%%%%%%%%% Merge with supplemental materials %%%%%%%%%%

\pagebreak
    %\widetext
    \onecolumngrid
    \clearpage
    \begin{center}
    \textbf{\large Supplemental Materials: Infinite temperature transport in the strong coupling regime of a nonintegrable quantum spin chain}
    \end{center}
    %%%%%%%%%% Merge with supplemental materials %%%%%%%%%%
    %%%%%%%%%% Prefix a "S" to all equations, figures, tables and reset the counter %%%%%%%%%%
    \setcounter{equation}{0}
    \setcounter{figure}{0}
    \setcounter{table}{0}
    \setcounter{section}{0}    
    \setcounter{page}{1}
    
    \makeatletter
    
    \renewcommand{\theequation}{S\arabic{equation}}

    \renewcommand{\thefigure}{S\arabic{figure}}
    \renewcommand{\theHfigure}{S\arabic{figure}}
    
    \renewcommand{\thetable}{S\arabic{table}}

    \setcounter{secnumdepth}{4}

    \renewcommand{\thesection}{S-\Roman{section}}    
    
    %%%%%%%%%% Prefix a "S" to all equations, figures, tables and reset the counter %%%%%%%%%%

    % FORCE NOT generate bibliography (in Arxiv it may duplicate bibliography,
    % use following lines to disable it) !!!
    \renewcommand{\bibliography}[1]{}
    \renewcommand{\bibliographystyle}[1]{}

\section{numerical details in calculating the spin conductivity}
    \label{sec:app:conductivity}

\subsection{exploiting symmetry}   \label{sec:supp:numerics:symm}
    \begin{figure}[H]
          \centering
          \scalebox{0.55}[0.55]{\includegraphics{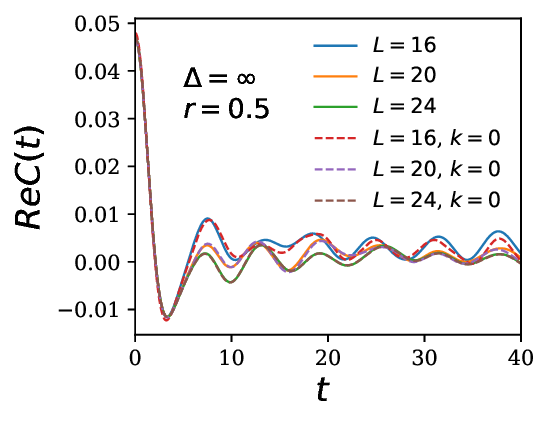}}
          \caption{\label{fig:supp:use:symm} 
          Comparison of current autocorrelation functions obtained
          with (solid lines) and without (dashed lines) average taken in all $k$
          sectors at $\Delta=\infty$ and $r=0.5$. For $L$ as large as 24, the
          difference between the two cases becomes indiscernible up to $t=40$. 
          }
          \end{figure}
    
    To extract the conductivity, first the current autocorrelation function needs to
    be evaluated numerically. Using conservation of magnetization and lattice
    momentum, the function $C_{m,k}(t)$ is evaluated for each $m$ and $k$
    sectors. Then ensemble average is taken to yield
    $C(t)=\frac{1}{2^L}\sum_{m,k}{\mathcal{N}_{m,k}}C_{m,k}(t)$, where
    $\mathcal{N}_{m,k}$ is the dimension of a sector labeled by $(m,k)$. In practice, for
    sufficiently large $L$, $C_{m,k}(t)$ barely depends on $k$, and making
    average over the $k$ sectors is not necessary (see also Ref.
    \cite{steinigeweg2015spin_supp}). An example for this is illustrated in
    Fig.\ref{fig:supp:use:symm}, and we confirm that this is also true when
    $L\ge 24$ for all other cases. As such, the results in the main text are all
    obtained in the $k=0$ sector with $24\le L \le34$.

\subsection{choice of $t_{max}$}    \label{sec:supp:numerics:tmax}
    \begin{figure}
          \centering
          %\scalebox{0.55}[0.55]{\includegraphics{supp_use_symm.eps}}
          \subfigure{\scalebox{0.55}[0.55]{\includegraphics[width=300mm]{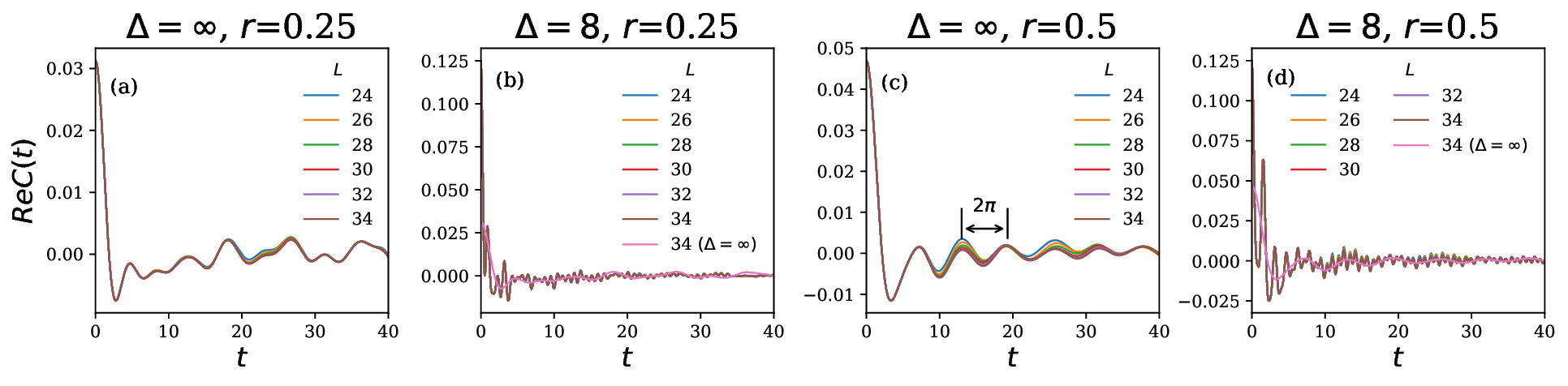}}}
          \subfigure{\scalebox{0.55}[0.55]{\includegraphics[width=300mm]{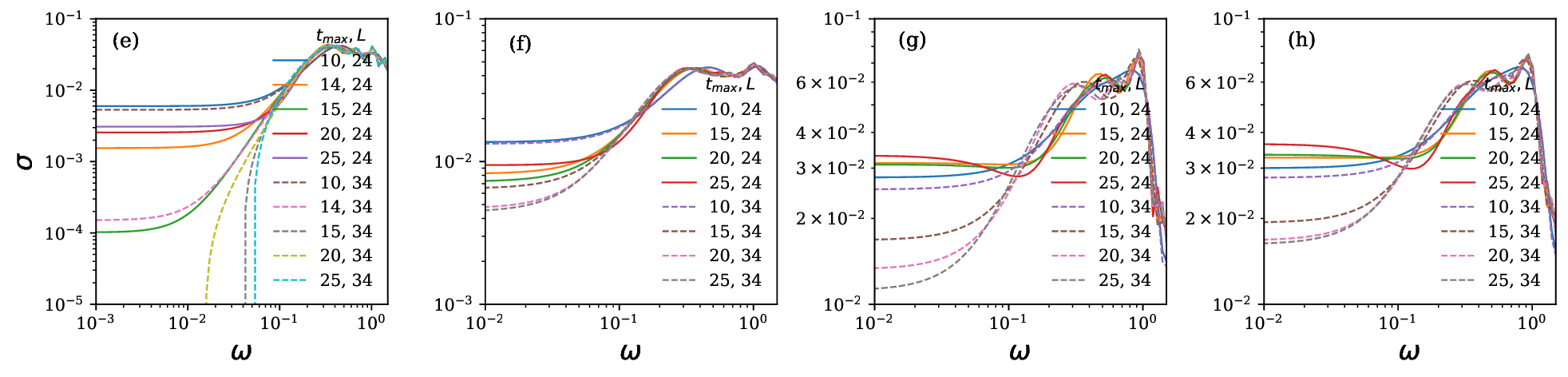}}}
          \caption{\label{fig:supp:tmax} 
            Top panels: Current correlation function for several $(\Delta,r)$
            pairs and different system sizes.   In panels (b) and (d), the results for
            $\Delta=\infty$ are also included for comparison.  Bottom panels:
            Low-frequency conductivity for the $(\Delta,r)$ pairs. It is evaluated by using
            different $t_{max}$ for the smallest and largest system sizes. 
          }
          \end{figure}

    Fig.\ref{fig:supp:tmax} shows the correlator and the conductivity for
    several typical $(\Delta,r)$ pairs. Although $C(t)$ can be accurately
    calculated for very long time using the DQT method, a cutoff time $t_{max}$
    is necessary due to restriction in system sizes. Our rule of thumb for
    determining an appropriate $t_{max}$   is to, on the one hand, extend
    $t_{max}$  as large as possible, on the other hand, avoid obviously 
    unreasonable results. By the latter we mean, for example, the negative
    conductivity obtained for $(\Delta,r)=(\infty, 0.25)$ [and also the
    irregular oscillations in $C(t)$] when \tmax $>14$; the fluctuation of the
    conductivity at low frequencies for $(\Delta, r)=(8, 0.5)$ when \tmax $>20$.
    Therefore, for the simulation in the main text, at  $r=0.25$ we use \tmax=14
    irrespective of $L$ and $\Delta$, and at $r=0.5$ using \tmax=20 also for all
    cases. It is difficult to assess the error induced by the finite \tmax.  The
    validity of our results is supported by  consistency in the data itself,
     well matching with the analytical scalings, and  obedience  to the
    sum-rule [see Fig.\ref{fig:supp:peak:pos} below]. 
    
    In passing, we note two salient features in the correlation function $C(t)$,
    which may be commonly found in strong coupling models. First, it drops to a negative
    value from the initial value in a short time for all cases. 
    %This cancellation is the cause of very depressed low-frequency conductivity.
    Second, then it oscillates for very long times---should be \emph{infinite} long for
    the kinetically constrained models at $\Delta=\infty$. In particular, at
    $(\Delta, r)=(\infty,0.5)$, the oscillation has a period $\approx 2\pi$,
    which leads to a pronounced peak in the conductivity at $\omega\approx1$. This is,
    microscopically, due to constantly tunneling between the two
    nearest-neighbor resonant configurations 
    $1101\leftrightarrow 1011$. At finite $\Delta$, there are
    higher-frequency oscillations riding on the previous one [see
    Fig.\ref{fig:supp:tmax}(d)], which will lead to peaks in the conductivity
    at intermediate and high frequencies (see Fig.\ref{fig:supp:peak:pos} below).
    
    %For example, at $(\Delta,r)=(\infty, 0.5)$, the oscillation has a period
    %$\approx2\pi$, which leads to a peak in the conductivity at
    %$\omega\approx 1$. This is due to transition between the
    %nearest-neighbor resonance process of $1101\leftrightarrow 1011$.
    %At $(\Delta,r)=(8, 0.5)$, there is a smaller period $\approx
    %2\pi/\Delta$ riding the former period.  This is due to the
    %non-resonant processes like $111 \leftrightarrow 1101$, leading to a
    %peak of the conductivity at $\omega=\Delta$ (see also discussions
    %below).             
    
    %By passing, we note several common features therein: (i) The correlator soon
    %drops to negative values from the initial positive values.  Cancellation of
    %them leads to suppressed low-frequency conductivity.  Microscopically, the
    %negative values are caused by backscattering of the magnons by the bound
    %states.  (iii) There are oscillations persisting for long times. These
    %oscillation can be explained by various transition between bound states.
    %(iii) The finite size effect is not significant in all cases, since the
    %systems are all close to localization.

\section{conductivity for a single particle on a lattice with OBC} \label{sec:supp:single:particle:OBC}
    %%%
    At infinite $\Delta$ a magnon can be strictly confined between two bound
    states. This is locally equivalent to a single particle moving on a finite
    chain of length $l$ with OBC, whose Hamiltonian
    is $H =  - t\sum_{j = 1}^{l - 1} {c_j^\dagger {c_{j + 1}} + c_{j +
    1}^\dagger {c_j}} $. Following \cite{rigol2008drude}, we determine the
    conductivity for this single particle problem.
    %Following \cite{rigol2008drude}, consider a single particle moving on a finite chain of length $l$ with open
    %boundary conditions (OBC), the Hamiltonian being $H =  - t\sum_{j = 1}^{l -
    %1} {c_j^\dagger {c_{j + 1}} + c_{j + 1}^\dagger {c_j}} $.  
    The eigenstates of this model are
    $|k\rangle\equiv \sum_{j=1}^{l}{A\sin(kj) |j\rangle}$, where  
    $A=\sqrt{2/(l+1)}$, $k=n\pi/(l+1)$ and $n=1$, 2, \dots, $l$. Every eigenstate  
    has a definite parity, which is the same with the parity of $n$. The corresponding
    eigenvalues are $\varepsilon(k)=-2t\cos(k)$. The current operator for OBC is
    (here setting $\hbar=1$ as usual)
    $\hat J =  - it\sum_{i = 1}^{L - 1} {(c_i^\dagger {c_{i + 1}} - c_{i +
    1}^\dagger {c_i})} $. The matrix elements for the current operator are 
    \[\langle k'|\hat J|k\rangle = - 2itA^2\frac{{\sin (k)\sin
    (k')}}{{\cos (k) - \cos (k')}}, \]    
    if $|k\rangle$ and $|k'\rangle$ have different parities, and zero otherwise. 
    The real part of the conductivity at infinite temperature, via the spectrum
    representation, reads 
        \[T\sigma (\omega ) = \frac{\pi }{{lZ}}\sum\limits_{k',k} {|\langle
        k'|J|k\rangle {|^2}\delta (\omega  - {\omega _{k',k}})}, \]
    where $Z=l$ and $\omega_{k',k}= 2t(\cos k - \cos k')$.  Elementary 
     considerations lead to
    that the square  of the matrix element  reaches its
    maximum $\max(|\langle k'|J|k\rangle|^2)\approx {(4t/\pi  )^2}$, when $k=\pi/2$ and $|k-k'|=\pi/(L+1)$. 
    Then the conductivity has its peak located at 
    $\omega^*=\omega_{k',k}\approx t(k^2-k'^2) \approx t\pi^2/[2(l+1)] \propto
    1/l$, for $l\gg 1$, with the weight  $\approx {(4t/ l )^2/\pi} \propto
    1/l^2$. So we get $T\sigma(\omega^*)\sim\omega^{*2}$. 

\section{Further results on spin conductivity at finite $\Delta$}
    \label{sec:app:conductivity:further}

\subsection{intermediate and high-frequency conductivity}
    \label{sec:supp:sigma:high:freq}
     \begin{figure}
          \centering
          \scalebox{0.55}[0.55]{\includegraphics{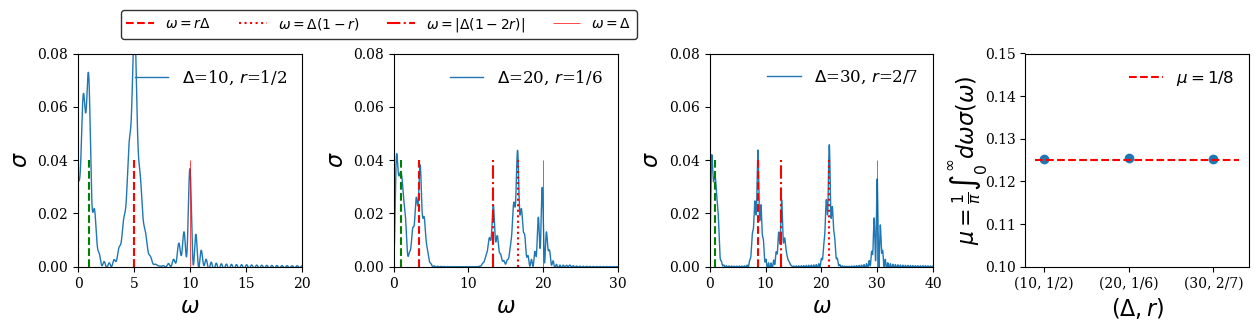}}
          \caption{\label{fig:supp:peak:pos} 
          Left three panels: Spin conductivity at three $(\Delta,r)$ pairs
          (as indicated in each panel) in the full frequency ranges. For each case, there is a peak at
          $\omega\approx 1$ (indicated by the dashed green line), the reason for
          which has been explained in the above. At intermediate and high frequencies ($\omega>1$),
          for each case, there are up to four peaks, whose positions are
          indicated in the legend on the top (at special $r$ values, say 0.5,
          the number of peaks is smaller than four). Right panel: The integrated conductivity
          $\mu=\frac{1}{\pi}\int_0^\infty{d\omega \sigma(\omega)}$ for the 
          $(\Delta,r)$ pairs. The sum-rule $\mu=\langle\hat J\hat
          J\rangle/L=1/8 $ is very well satisfied. 
          }
          \end{figure}
   
    Fig.\ref{fig:supp:peak:pos} shows conductivity for several $(\Delta,r)$
    pairs in their full frequency ranges. The most significant feature is that, for $\omega>1$,
    there are up to four pronounced peaks located at certain frequencies
    $\omega^*$, and the spectral weights are distributed approximately in
    $[\omega^*-2,\omega^*+2]$ for each $\omega^*$. 
    Microscopically, these peaks are caused by transition
    between different types of bound states. For all the transitions, there are 
    in total four distinct energy differences (leading to distinct $\omega^*$), as 
    exemplified by the following equation 
    \begin{equation}
    \begin{split}
         1101     &\leftrightarrow 11001  \quad\quad         r\Delta \\
         111001   &\leftrightarrow 110101 \quad\quad           \Delta(1-r)  \\
         10101    &\leftrightarrow 10011  \quad\quad          |\Delta(1-2r)| \\
         111      &\leftrightarrow 1101   \quad\quad\quad        \Delta. \\
    \end{split}
    \end{equation}
    The number and  positions of $\omega^*$ in this equation completely match with the numerical data. 
    In addition, the right panel of Fig.\ref{fig:supp:peak:pos} shows that the sum-rule is very
    well satisfied. 
    We also note that the intermediate and high frequency peaks is a
    common feature for the strongly coupling models. Similar results have been
    reported in \cite{maldague1977optical} for the Hubbard model many years ago, and also in
    \cite{sanchez2018anomalous}.

\subsection{low-frequency conductivity}
    \label{sec:supp:sigma:low:freq}
     \begin{figure}[H]
          \centering
          \scalebox{0.55}[0.55]{\includegraphics{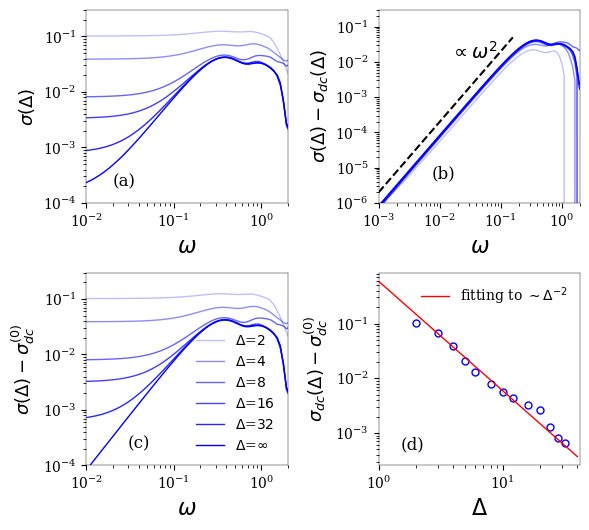}}
          \caption{\label{fig:supp:sigma:finite:delta} 
          Low-frequency conductivity at finite $\Delta$ and $r=0.25$. These
          results are obtained in the same way as that of
          Fig.\ref{fig:sigma:finite:V}, except that $r$ is changed from 0.5 to
          0.25 (see the caption of the previous figure). Note that at $r=0.25$ the
          quantity $\sigma_{dc}^{(0)}=0.00015$ is very small, so it makes little
          difference whether extracting it from $\sigma(\Delta)$ [subplot (a)] or not
          [subplot (c)].}
          \end{figure}

    In the main text we have shown that, at finite $\Delta$ and $r=0.5$, the
    finite-$\Delta$ correction to the dc
    conductivity scales as $\sim \Delta^{-2}$. Here we show supplemental results for
    $r=0.25$ in Fig.\ref{fig:supp:sigma:finite:delta}, in which the same scaling is obtained. This implies that this scaling
    should hold in the entire range of $0<r<1$ for sufficiently large $\Delta$.

%\begin{thebibliography}{11}
%\bibitem{S_RefA} A. Someone, C. Someone, D. Someone, Phys. Rev. Lett. {\bf 11}, 1111 (1911).
%\end{thebibliography}    

% 
    %https://tex.stackexchange.com/questions/168169/options-for-supplementary-materials-in-preprint-version-revtex-arxiv 
    
%\section*{References}
%    \bibliography{bib-for-supp-mat.bib}

\begin{thebibliography}{65}%
\makeatletter
\providecommand \@ifxundefined [1]{%
 \@ifx{#1\undefined}
}%
\providecommand \@ifnum [1]{%
 \ifnum #1\expandafter \@firstoftwo
 \else \expandafter \@secondoftwo
 \fi
}%
\providecommand \@ifx [1]{%
 \ifx #1\expandafter \@firstoftwo
 \else \expandafter \@secondoftwo
 \fi
}%
\providecommand \natexlab [1]{#1}%
\providecommand \enquote  [1]{``#1''}%
\providecommand \bibnamefont  [1]{#1}%
\providecommand \bibfnamefont [1]{#1}%
\providecommand \citenamefont [1]{#1}%
\providecommand \href@noop [0]{\@secondoftwo}%
\providecommand \href [0]{\begingroup \@sanitize@url \@href}%
\providecommand \@href[1]{\@@startlink{#1}\@@href}%
\providecommand \@@href[1]{\endgroup#1\@@endlink}%
\providecommand \@sanitize@url [0]{\catcode `\\12\catcode `\$12\catcode
  `\&12\catcode `\#12\catcode `\^12\catcode `\_12\catcode `\%12\relax}%
\providecommand \@@startlink[1]{}%
\providecommand \@@endlink[0]{}%
\providecommand \url  [0]{\begingroup\@sanitize@url \@url }%
\providecommand \@url [1]{\endgroup\@href {#1}{\urlprefix }}%
\providecommand \urlprefix  [0]{URL }%
\providecommand \Eprint [0]{\href }%
\providecommand \doibase [0]{https://doi.org/}%
\providecommand \selectlanguage [0]{\@gobble}%
\providecommand \bibinfo  [0]{\@secondoftwo}%
\providecommand \bibfield  [0]{\@secondoftwo}%
\providecommand \translation [1]{[#1]}%
\providecommand \BibitemOpen [0]{}%
\providecommand \bibitemStop [0]{}%
\providecommand \bibitemNoStop [0]{.\EOS\space}%
\providecommand \EOS [0]{\spacefactor3000\relax}%
\providecommand \BibitemShut  [1]{\csname bibitem#1\endcsname}%
\let\auto@bib@innerbib\@empty
%</preamble>
\bibitem [{\citenamefont {Bertini}\ \emph {et~al.}(2021)\citenamefont
  {Bertini}, \citenamefont {Heidrich-Meisner}, \citenamefont {Karrasch},
  \citenamefont {Prosen}, \citenamefont {Steinigeweg},\ and\ \citenamefont
  {\ifmmode \check{Z}\else \v{Z}\fi{}nidari\ifmmode~\check{c}\else
  \v{c}\fi{}}}]{bertini2020finite}%
  \BibitemOpen
  \bibfield  {author} {\bibinfo {author} {\bibfnamefont {B.}~\bibnamefont
  {Bertini}}, \bibinfo {author} {\bibfnamefont {F.}~\bibnamefont
  {Heidrich-Meisner}}, \bibinfo {author} {\bibfnamefont {C.}~\bibnamefont
  {Karrasch}}, \bibinfo {author} {\bibfnamefont {T.}~\bibnamefont {Prosen}},
  \bibinfo {author} {\bibfnamefont {R.}~\bibnamefont {Steinigeweg}},\ and\
  \bibinfo {author} {\bibfnamefont {M.}~\bibnamefont {\ifmmode \check{Z}\else
  \v{Z}\fi{}nidari\ifmmode~\check{c}\else \v{c}\fi{}}},\ }\bibfield  {title}
  {\bibinfo {title} {Finite-temperature transport in one-dimensional quantum
  lattice models},\ }\href {https://doi.org/10.1103/RevModPhys.93.025003}
  {\bibfield  {journal} {\bibinfo  {journal} {Rev. Mod. Phys.}\ }\textbf
  {\bibinfo {volume} {93}},\ \bibinfo {pages} {025003} (\bibinfo {year}
  {2021})}\BibitemShut {NoStop}%
\bibitem [{\citenamefont {Castella}\ \emph {et~al.}(1995)\citenamefont
  {Castella}, \citenamefont {Zotos},\ and\ \citenamefont
  {Prelovšek}}]{castella1995integrability}%
  \BibitemOpen
  \bibfield  {author} {\bibinfo {author} {\bibfnamefont {H.}~\bibnamefont
  {Castella}}, \bibinfo {author} {\bibfnamefont {X.}~\bibnamefont {Zotos}},\
  and\ \bibinfo {author} {\bibfnamefont {P.}~\bibnamefont {Prelovšek}},\
  }\bibfield  {title} {\bibinfo {title} {Integrability and {Ideal}
  {Conductance} at {Finite} {Temperatures}},\ }\href
  {https://doi.org/10.1103/PhysRevLett.74.972} {\bibfield  {journal} {\bibinfo
  {journal} {Phys. Rev. Lett.}\ }\textbf {\bibinfo {volume} {74}},\ \bibinfo
  {pages} {972} (\bibinfo {year} {1995})}\BibitemShut {NoStop}%
\bibitem [{\citenamefont {Zotos}\ and\ \citenamefont
  {Prelovšek}(1996)}]{zotos1996evidence}%
  \BibitemOpen
  \bibfield  {author} {\bibinfo {author} {\bibfnamefont {X.}~\bibnamefont
  {Zotos}}\ and\ \bibinfo {author} {\bibfnamefont {P.}~\bibnamefont
  {Prelovšek}},\ }\bibfield  {title} {\bibinfo {title} {Evidence for ideal
  insulating or conducting state in a one-dimensional integrable system},\
  }\href {https://doi.org/10.1103/PhysRevB.53.983} {\bibfield  {journal}
  {\bibinfo  {journal} {Phys. Rev. B}\ }\textbf {\bibinfo {volume} {53}},\
  \bibinfo {pages} {983} (\bibinfo {year} {1996})}\BibitemShut {NoStop}%
\bibitem [{\citenamefont {Karrasch}\ \emph {et~al.}(2013)\citenamefont
  {Karrasch}, \citenamefont {Hauschild}, \citenamefont {Langer},\ and\
  \citenamefont {Heidrich-Meisner}}]{karrasch2013drude}%
  \BibitemOpen
  \bibfield  {author} {\bibinfo {author} {\bibfnamefont {C.}~\bibnamefont
  {Karrasch}}, \bibinfo {author} {\bibfnamefont {J.}~\bibnamefont {Hauschild}},
  \bibinfo {author} {\bibfnamefont {S.}~\bibnamefont {Langer}},\ and\ \bibinfo
  {author} {\bibfnamefont {F.}~\bibnamefont {Heidrich-Meisner}},\ }\bibfield
  {title} {\bibinfo {title} {Drude weight of the
  spin-\${\textbackslash}frac\{1\}\{2\}\$ {XXZ} chain: {Density} matrix
  renormalization group versus exact diagonalization},\ }\href
  {https://doi.org/10.1103/PhysRevB.87.245128} {\bibfield  {journal} {\bibinfo
  {journal} {Phys. Rev. B}\ }\textbf {\bibinfo {volume} {87}},\ \bibinfo
  {pages} {245128} (\bibinfo {year} {2013})}\BibitemShut {NoStop}%
\bibitem [{\citenamefont {Karrasch}(2017)}]{karrasch2017hubbard}%
  \BibitemOpen
  \bibfield  {author} {\bibinfo {author} {\bibfnamefont {C.}~\bibnamefont
  {Karrasch}},\ }\bibfield  {title} {\bibinfo {title} {Hubbard-to-{Heisenberg}
  crossover (and efficient computation) of {Drude} weights at low
  temperatures},\ }\href {https://doi.org/10.1088/1367-2630/aa631a} {\bibfield
  {journal} {\bibinfo  {journal} {New J. Phys.}\ }\textbf {\bibinfo {volume}
  {19}},\ \bibinfo {pages} {033027} (\bibinfo {year} {2017})}\BibitemShut
  {NoStop}%
\bibitem [{\citenamefont {Ljubotina}\ \emph {et~al.}(2019)\citenamefont
  {Ljubotina}, \citenamefont {Žnidarič},\ and\ \citenamefont
  {Prosen}}]{ljubotina2019kardar}%
  \BibitemOpen
  \bibfield  {author} {\bibinfo {author} {\bibfnamefont {M.}~\bibnamefont
  {Ljubotina}}, \bibinfo {author} {\bibfnamefont {M.}~\bibnamefont
  {Žnidarič}},\ and\ \bibinfo {author} {\bibfnamefont {T.}~\bibnamefont
  {Prosen}},\ }\bibfield  {title} {\bibinfo {title} {Kardar-{Parisi}-{Zhang}
  {Physics} in the {Quantum} {Heisenberg} {Magnet}},\ }\href
  {https://doi.org/10.1103/PhysRevLett.122.210602} {\bibfield  {journal}
  {\bibinfo  {journal} {Phys. Rev. Lett.}\ }\textbf {\bibinfo {volume} {122}},\
  \bibinfo {pages} {210602} (\bibinfo {year} {2019})}\BibitemShut {NoStop}%
\bibitem [{\citenamefont {Prosen}\ and\ \citenamefont
  {Ilievski}(2013)}]{prosen2013families}%
  \BibitemOpen
  \bibfield  {author} {\bibinfo {author} {\bibfnamefont {T.}~\bibnamefont
  {Prosen}}\ and\ \bibinfo {author} {\bibfnamefont {E.}~\bibnamefont
  {Ilievski}},\ }\bibfield  {title} {\bibinfo {title} {Families of {Quasilocal}
  {Conservation} {Laws} and {Quantum} {Spin} {Transport}},\ }\href
  {https://doi.org/10.1103/PhysRevLett.111.057203} {\bibfield  {journal}
  {\bibinfo  {journal} {Phys. Rev. Lett.}\ }\textbf {\bibinfo {volume} {111}},\
  \bibinfo {pages} {057203} (\bibinfo {year} {2013})}\BibitemShut {NoStop}%
\bibitem [{\citenamefont {Bertini}\ \emph {et~al.}(2016)\citenamefont
  {Bertini}, \citenamefont {Collura}, \citenamefont {De~Nardis},\ and\
  \citenamefont {Fagotti}}]{bertini2016transport}%
  \BibitemOpen
  \bibfield  {author} {\bibinfo {author} {\bibfnamefont {B.}~\bibnamefont
  {Bertini}}, \bibinfo {author} {\bibfnamefont {M.}~\bibnamefont {Collura}},
  \bibinfo {author} {\bibfnamefont {J.}~\bibnamefont {De~Nardis}},\ and\
  \bibinfo {author} {\bibfnamefont {M.}~\bibnamefont {Fagotti}},\ }\bibfield
  {title} {\bibinfo {title} {Transport in {Out}-of-{Equilibrium} \${XXZ}\$
  {Chains}: {Exact} {Profiles} of {Charges} and {Currents}},\ }\href
  {https://doi.org/10.1103/PhysRevLett.117.207201} {\bibfield  {journal}
  {\bibinfo  {journal} {Phys. Rev. Lett.}\ }\textbf {\bibinfo {volume} {117}},\
  \bibinfo {pages} {207201} (\bibinfo {year} {2016})}\BibitemShut {NoStop}%
\bibitem [{\citenamefont {Castro-Alvaredo}\ \emph {et~al.}(2016)\citenamefont
  {Castro-Alvaredo}, \citenamefont {Doyon},\ and\ \citenamefont
  {Yoshimura}}]{castroalvaredo2016emergent}%
  \BibitemOpen
  \bibfield  {author} {\bibinfo {author} {\bibfnamefont {O.~A.}\ \bibnamefont
  {Castro-Alvaredo}}, \bibinfo {author} {\bibfnamefont {B.}~\bibnamefont
  {Doyon}},\ and\ \bibinfo {author} {\bibfnamefont {T.}~\bibnamefont
  {Yoshimura}},\ }\bibfield  {title} {\bibinfo {title} {Emergent
  {Hydrodynamics} in {Integrable} {Quantum} {Systems} {Out} of {Equilibrium}},\
  }\href {https://doi.org/10.1103/PhysRevX.6.041065} {\bibfield  {journal}
  {\bibinfo  {journal} {Phys. Rev. X}\ }\textbf {\bibinfo {volume} {6}},\
  \bibinfo {pages} {041065} (\bibinfo {year} {2016})}\BibitemShut {NoStop}%
\bibitem [{\citenamefont {Ilievski}\ and\ \citenamefont
  {De~Nardis}(2017)}]{ilievski2017microscopic}%
  \BibitemOpen
  \bibfield  {author} {\bibinfo {author} {\bibfnamefont {E.}~\bibnamefont
  {Ilievski}}\ and\ \bibinfo {author} {\bibfnamefont {J.}~\bibnamefont
  {De~Nardis}},\ }\bibfield  {title} {\bibinfo {title} {Microscopic {Origin} of
  {Ideal} {Conductivity} in {Integrable} {Quantum} {Models}},\ }\href
  {https://doi.org/10.1103/PhysRevLett.119.020602} {\bibfield  {journal}
  {\bibinfo  {journal} {Phys. Rev. Lett.}\ }\textbf {\bibinfo {volume} {119}},\
  \bibinfo {pages} {020602} (\bibinfo {year} {2017})}\BibitemShut {NoStop}%
\bibitem [{\citenamefont {De~Nardis}\ \emph {et~al.}(2018)\citenamefont
  {De~Nardis}, \citenamefont {Bernard},\ and\ \citenamefont
  {Doyon}}]{denardis2018hydrodynamic}%
  \BibitemOpen
  \bibfield  {author} {\bibinfo {author} {\bibfnamefont {J.}~\bibnamefont
  {De~Nardis}}, \bibinfo {author} {\bibfnamefont {D.}~\bibnamefont {Bernard}},\
  and\ \bibinfo {author} {\bibfnamefont {B.}~\bibnamefont {Doyon}},\ }\bibfield
   {title} {\bibinfo {title} {Hydrodynamic {Diffusion} in {Integrable}
  {Systems}},\ }\href {https://doi.org/10.1103/PhysRevLett.121.160603}
  {\bibfield  {journal} {\bibinfo  {journal} {Phys. Rev. Lett.}\ }\textbf
  {\bibinfo {volume} {121}},\ \bibinfo {pages} {160603} (\bibinfo {year}
  {2018})}\BibitemShut {NoStop}%
\bibitem [{\citenamefont {Bulchandani}\ \emph {et~al.}(2021)\citenamefont
  {Bulchandani}, \citenamefont {Gopalakrishnan},\ and\ \citenamefont
  {Ilievski}}]{bulchandani2021superdiffusion}%
  \BibitemOpen
  \bibfield  {author} {\bibinfo {author} {\bibfnamefont {V.~B.}\ \bibnamefont
  {Bulchandani}}, \bibinfo {author} {\bibfnamefont {S.}~\bibnamefont
  {Gopalakrishnan}},\ and\ \bibinfo {author} {\bibfnamefont {E.}~\bibnamefont
  {Ilievski}},\ }\bibfield  {title} {\bibinfo {title} {Superdiffusion in spin
  chains},\ }\href {https://doi.org/10.1088/1742-5468/ac12c7} {\bibfield
  {journal} {\bibinfo  {journal} {J. Stat. Mech.}\ }\textbf {\bibinfo {volume}
  {2021}},\ \bibinfo {pages} {084001} (\bibinfo {year} {2021})}\BibitemShut
  {NoStop}%
\bibitem [{\citenamefont {Koster}\ and\ \citenamefont
  {Slater}(1954)}]{koster1954wave}%
  \BibitemOpen
  \bibfield  {author} {\bibinfo {author} {\bibfnamefont {G.~F.}\ \bibnamefont
  {Koster}}\ and\ \bibinfo {author} {\bibfnamefont {J.~C.}\ \bibnamefont
  {Slater}},\ }\bibfield  {title} {\bibinfo {title} {Wave {Functions} for
  {Impurity} {Levels}},\ }\href {https://doi.org/10.1103/PhysRev.95.1167}
  {\bibfield  {journal} {\bibinfo  {journal} {Phys. Rev.}\ }\textbf {\bibinfo
  {volume} {95}},\ \bibinfo {pages} {1167} (\bibinfo {year}
  {1954})}\BibitemShut {NoStop}%
\bibitem [{\citenamefont {Kagan}\ and\ \citenamefont
  {Maksimov}(1984)}]{kagan1984localization}%
  \BibitemOpen
  \bibfield  {author} {\bibinfo {author} {\bibfnamefont {Y.}~\bibnamefont
  {Kagan}}\ and\ \bibinfo {author} {\bibfnamefont {L.~A.}\ \bibnamefont
  {Maksimov}},\ }\bibfield  {title} {\bibinfo {title} {Localization in a system
  of interacting particles diffusing in a regular crystal},\ }\href
  {https://ui.adsabs.harvard.edu/abs/1984JETP...60..201K} {\bibfield  {journal}
  {\bibinfo  {journal} {J. Exp. Theor. Phys.}\ }\textbf {\bibinfo {volume}
  {60}},\ \bibinfo {pages} {201} (\bibinfo {year} {1984})}\BibitemShut
  {NoStop}%
\bibitem [{\citenamefont {Mattis}(1986)}]{mattis1986few}%
  \BibitemOpen
  \bibfield  {author} {\bibinfo {author} {\bibfnamefont {D.~C.}\ \bibnamefont
  {Mattis}},\ }\bibfield  {title} {\bibinfo {title} {The few-body problem on a
  lattice},\ }\href {https://doi.org/10.1103/RevModPhys.58.361} {\bibfield
  {journal} {\bibinfo  {journal} {Rev. Mod. Phys.}\ }\textbf {\bibinfo {volume}
  {58}},\ \bibinfo {pages} {361} (\bibinfo {year} {1986})}\BibitemShut
  {NoStop}%
\bibitem [{\citenamefont {Winkler}\ \emph {et~al.}(2006)\citenamefont
  {Winkler}, \citenamefont {Thalhammer}, \citenamefont {Lang}, \citenamefont
  {Grimm}, \citenamefont {Hecker~Denschlag}, \citenamefont {Daley},
  \citenamefont {Kantian}, \citenamefont {Büchler},\ and\ \citenamefont
  {Zoller}}]{winkler2006repulsively}%
  \BibitemOpen
  \bibfield  {author} {\bibinfo {author} {\bibfnamefont {K.}~\bibnamefont
  {Winkler}}, \bibinfo {author} {\bibfnamefont {G.}~\bibnamefont {Thalhammer}},
  \bibinfo {author} {\bibfnamefont {F.}~\bibnamefont {Lang}}, \bibinfo {author}
  {\bibfnamefont {R.}~\bibnamefont {Grimm}}, \bibinfo {author} {\bibfnamefont
  {J.}~\bibnamefont {Hecker~Denschlag}}, \bibinfo {author} {\bibfnamefont
  {A.~J.}\ \bibnamefont {Daley}}, \bibinfo {author} {\bibfnamefont
  {A.}~\bibnamefont {Kantian}}, \bibinfo {author} {\bibfnamefont {H.~P.}\
  \bibnamefont {Büchler}},\ and\ \bibinfo {author} {\bibfnamefont
  {P.}~\bibnamefont {Zoller}},\ }\bibfield  {title} {\bibinfo {title}
  {Repulsively bound atom pairs in an optical lattice},\ }\href
  {https://doi.org/10.1038/nature04918} {\bibfield  {journal} {\bibinfo
  {journal} {Nature}\ }\textbf {\bibinfo {volume} {441}},\ \bibinfo {pages}
  {853} (\bibinfo {year} {2006})}\BibitemShut {NoStop}%
\bibitem [{\citenamefont {Valiente}\ \emph {et~al.}(2010)\citenamefont
  {Valiente}, \citenamefont {Petrosyan},\ and\ \citenamefont
  {Saenz}}]{valiente2010three}%
  \BibitemOpen
  \bibfield  {author} {\bibinfo {author} {\bibfnamefont {M.}~\bibnamefont
  {Valiente}}, \bibinfo {author} {\bibfnamefont {D.}~\bibnamefont
  {Petrosyan}},\ and\ \bibinfo {author} {\bibfnamefont {A.}~\bibnamefont
  {Saenz}},\ }\bibfield  {title} {\bibinfo {title} {Three-body bound states in
  a lattice},\ }\href {https://doi.org/10.1103/PhysRevA.81.011601} {\bibfield
  {journal} {\bibinfo  {journal} {Phys. Rev. A}\ }\textbf {\bibinfo {volume}
  {81}},\ \bibinfo {pages} {011601} (\bibinfo {year} {2010})}\BibitemShut
  {NoStop}%
\bibitem [{\citenamefont {Fukuhara}\ \emph {et~al.}(2013)\citenamefont
  {Fukuhara}, \citenamefont {Schauß}, \citenamefont {Endres}, \citenamefont
  {Hild}, \citenamefont {Cheneau}, \citenamefont {Bloch},\ and\ \citenamefont
  {Gross}}]{fukuhara2013microscopic}%
  \BibitemOpen
  \bibfield  {author} {\bibinfo {author} {\bibfnamefont {T.}~\bibnamefont
  {Fukuhara}}, \bibinfo {author} {\bibfnamefont {P.}~\bibnamefont {Schauß}},
  \bibinfo {author} {\bibfnamefont {M.}~\bibnamefont {Endres}}, \bibinfo
  {author} {\bibfnamefont {S.}~\bibnamefont {Hild}}, \bibinfo {author}
  {\bibfnamefont {M.}~\bibnamefont {Cheneau}}, \bibinfo {author} {\bibfnamefont
  {I.}~\bibnamefont {Bloch}},\ and\ \bibinfo {author} {\bibfnamefont
  {C.}~\bibnamefont {Gross}},\ }\bibfield  {title} {\bibinfo {title}
  {Microscopic observation of magnon bound states and their dynamics},\ }\href
  {https://doi.org/10.1038/nature12541} {\bibfield  {journal} {\bibinfo
  {journal} {Nature}\ }\textbf {\bibinfo {volume} {502}},\ \bibinfo {pages}
  {76} (\bibinfo {year} {2013})}\BibitemShut {NoStop}%
\bibitem [{\citenamefont {Kranzl}\ \emph {et~al.}(2023)\citenamefont {Kranzl},
  \citenamefont {Birnkammer}, \citenamefont {Joshi}, \citenamefont
  {Bastianello}, \citenamefont {Blatt}, \citenamefont {Knap},\ and\
  \citenamefont {Roos}}]{kranzl2023observation}%
  \BibitemOpen
  \bibfield  {author} {\bibinfo {author} {\bibfnamefont {F.}~\bibnamefont
  {Kranzl}}, \bibinfo {author} {\bibfnamefont {S.}~\bibnamefont {Birnkammer}},
  \bibinfo {author} {\bibfnamefont {M.~K.}\ \bibnamefont {Joshi}}, \bibinfo
  {author} {\bibfnamefont {A.}~\bibnamefont {Bastianello}}, \bibinfo {author}
  {\bibfnamefont {R.}~\bibnamefont {Blatt}}, \bibinfo {author} {\bibfnamefont
  {M.}~\bibnamefont {Knap}},\ and\ \bibinfo {author} {\bibfnamefont {C.~F.}\
  \bibnamefont {Roos}},\ }\bibfield  {title} {\bibinfo {title} {Observation of
  {Magnon} {Bound} {States} in the {Long}-{Range}, {Anisotropic} {Heisenberg}
  {Model}},\ }\href {https://doi.org/10.1103/PhysRevX.13.031017} {\bibfield
  {journal} {\bibinfo  {journal} {Phys. Rev. X}\ }\textbf {\bibinfo {volume}
  {13}},\ \bibinfo {pages} {031017} (\bibinfo {year} {2023})}\BibitemShut
  {NoStop}%
\bibitem [{\citenamefont {Schmiedinghoff}\ \emph {et~al.}(2022)\citenamefont
  {Schmiedinghoff}, \citenamefont {Müller}, \citenamefont {Kumar},
  \citenamefont {Uhrig},\ and\ \citenamefont
  {Fauseweh}}]{schmiedinghoff2022three}%
  \BibitemOpen
  \bibfield  {author} {\bibinfo {author} {\bibfnamefont {G.}~\bibnamefont
  {Schmiedinghoff}}, \bibinfo {author} {\bibfnamefont {L.}~\bibnamefont
  {Müller}}, \bibinfo {author} {\bibfnamefont {U.}~\bibnamefont {Kumar}},
  \bibinfo {author} {\bibfnamefont {G.~S.}\ \bibnamefont {Uhrig}},\ and\
  \bibinfo {author} {\bibfnamefont {B.}~\bibnamefont {Fauseweh}},\ }\bibfield
  {title} {\bibinfo {title} {Three-body bound states in antiferromagnetic spin
  ladders},\ }\href {https://doi.org/10.1038/s42005-022-00986-0} {\bibfield
  {journal} {\bibinfo  {journal} {Commun Phys}\ }\textbf {\bibinfo {volume}
  {5}},\ \bibinfo {pages} {1} (\bibinfo {year} {2022})}\BibitemShut {NoStop}%
\bibitem [{\citenamefont {Li}(2023)}]{li2023slow}%
  \BibitemOpen
  \bibfield  {author} {\bibinfo {author} {\bibfnamefont {Z.-H.}\ \bibnamefont
  {Li}},\ }\bibfield  {title} {\bibinfo {title} {Slow transport and bound
  states for spinless fermions with long-range {Coulomb} interactions on
  one-dimensional lattices},\ }\href
  {https://doi.org/10.1103/PhysRevB.108.045131} {\bibfield  {journal} {\bibinfo
   {journal} {Phys. Rev. B}\ }\textbf {\bibinfo {volume} {108}},\ \bibinfo
  {pages} {045131} (\bibinfo {year} {2023})}\BibitemShut {NoStop}%
\bibitem [{\citenamefont {De~Roeck}\ and\ \citenamefont
  {Huveneers}(2014)}]{roeck2014asymptotic}%
  \BibitemOpen
  \bibfield  {author} {\bibinfo {author} {\bibfnamefont {W.}~\bibnamefont
  {De~Roeck}}\ and\ \bibinfo {author} {\bibfnamefont {F.}~\bibnamefont
  {Huveneers}},\ }\bibfield  {title} {\bibinfo {title} {Asymptotic {Quantum}
  {Many}-{Body} {Localization} from {Thermal} {Disorder}},\ }\href
  {https://doi.org/10.1007/s00220-014-2116-8} {\bibfield  {journal} {\bibinfo
  {journal} {Commun. Math. Phys.}\ }\textbf {\bibinfo {volume} {332}},\
  \bibinfo {pages} {1017} (\bibinfo {year} {2014})}\BibitemShut {NoStop}%
\bibitem [{\citenamefont {Schiulaz}\ and\ \citenamefont
  {Müller}(2014)}]{schiulaz2014ideal}%
  \BibitemOpen
  \bibfield  {author} {\bibinfo {author} {\bibfnamefont {M.}~\bibnamefont
  {Schiulaz}}\ and\ \bibinfo {author} {\bibfnamefont {M.}~\bibnamefont
  {Müller}},\ }\bibfield  {title} {\bibinfo {title} {Ideal quantum glass
  transitions: {Many}-body localization without quenched disorder},\ }\href
  {https://doi.org/10.1063/1.4893505} {\bibfield  {journal} {\bibinfo
  {journal} {AIP Conf. Proc.}\ }\textbf {\bibinfo {volume} {1610}},\ \bibinfo
  {pages} {11} (\bibinfo {year} {2014})}\BibitemShut {NoStop}%
\bibitem [{\citenamefont {Schiulaz}\ \emph {et~al.}(2015)\citenamefont
  {Schiulaz}, \citenamefont {Silva},\ and\ \citenamefont
  {Müller}}]{schiulaz2015dynamics}%
  \BibitemOpen
  \bibfield  {author} {\bibinfo {author} {\bibfnamefont {M.}~\bibnamefont
  {Schiulaz}}, \bibinfo {author} {\bibfnamefont {A.}~\bibnamefont {Silva}},\
  and\ \bibinfo {author} {\bibfnamefont {M.}~\bibnamefont {Müller}},\
  }\bibfield  {title} {\bibinfo {title} {Dynamics in many-body localized
  quantum systems without disorder},\ }\href
  {https://doi.org/10.1103/PhysRevB.91.184202} {\bibfield  {journal} {\bibinfo
  {journal} {Phys. Rev. B}\ }\textbf {\bibinfo {volume} {91}},\ \bibinfo
  {pages} {184202} (\bibinfo {year} {2015})}\BibitemShut {NoStop}%
\bibitem [{\citenamefont {Yao}\ \emph {et~al.}(2016)\citenamefont {Yao},
  \citenamefont {Laumann}, \citenamefont {Cirac}, \citenamefont {Lukin},\ and\
  \citenamefont {Moore}}]{yao2016quasi}%
  \BibitemOpen
  \bibfield  {author} {\bibinfo {author} {\bibfnamefont {N.}~\bibnamefont
  {Yao}}, \bibinfo {author} {\bibfnamefont {C.}~\bibnamefont {Laumann}},
  \bibinfo {author} {\bibfnamefont {J.}~\bibnamefont {Cirac}}, \bibinfo
  {author} {\bibfnamefont {M.}~\bibnamefont {Lukin}},\ and\ \bibinfo {author}
  {\bibfnamefont {J.}~\bibnamefont {Moore}},\ }\bibfield  {title} {\bibinfo
  {title} {Quasi-{Many}-{Body} {Localization} in {Translation}-{Invariant}
  {Systems}},\ }\href {https://doi.org/10.1103/PhysRevLett.117.240601}
  {\bibfield  {journal} {\bibinfo  {journal} {Phys. Rev. Lett.}\ }\textbf
  {\bibinfo {volume} {117}},\ \bibinfo {pages} {240601} (\bibinfo {year}
  {2016})}\BibitemShut {NoStop}%
\bibitem [{\citenamefont {Barbiero}\ \emph {et~al.}(2015)\citenamefont
  {Barbiero}, \citenamefont {Menotti}, \citenamefont {Recati},\ and\
  \citenamefont {Santos}}]{barbiero2015out}%
  \BibitemOpen
  \bibfield  {author} {\bibinfo {author} {\bibfnamefont {L.}~\bibnamefont
  {Barbiero}}, \bibinfo {author} {\bibfnamefont {C.}~\bibnamefont {Menotti}},
  \bibinfo {author} {\bibfnamefont {A.}~\bibnamefont {Recati}},\ and\ \bibinfo
  {author} {\bibfnamefont {L.}~\bibnamefont {Santos}},\ }\bibfield  {title}
  {\bibinfo {title} {Out-of-equilibrium states and quasi-many-body localization
  in polar lattice gases},\ }\href {https://doi.org/10.1103/PhysRevB.92.180406}
  {\bibfield  {journal} {\bibinfo  {journal} {Phys. Rev. B}\ }\textbf {\bibinfo
  {volume} {92}},\ \bibinfo {pages} {180406(R)} (\bibinfo {year}
  {2015})}\BibitemShut {NoStop}%
\bibitem [{\citenamefont {Abanin}\ \emph {et~al.}(2017)\citenamefont {Abanin},
  \citenamefont {De~Roeck}, \citenamefont {Ho},\ and\ \citenamefont
  {Huveneers}}]{abanin2017rigorous}%
  \BibitemOpen
  \bibfield  {author} {\bibinfo {author} {\bibfnamefont {D.}~\bibnamefont
  {Abanin}}, \bibinfo {author} {\bibfnamefont {W.}~\bibnamefont {De~Roeck}},
  \bibinfo {author} {\bibfnamefont {W.~W.}\ \bibnamefont {Ho}},\ and\ \bibinfo
  {author} {\bibfnamefont {F.}~\bibnamefont {Huveneers}},\ }\bibfield  {title}
  {\bibinfo {title} {A {Rigorous} {Theory} of {Many}-{Body} {Prethermalization}
  for {Periodically} {Driven} and {Closed} {Quantum} {Systems}},\ }\href
  {https://doi.org/10.1007/s00220-017-2930-x} {\bibfield  {journal} {\bibinfo
  {journal} {Commun. Math. Phys.}\ }\textbf {\bibinfo {volume} {354}},\
  \bibinfo {pages} {809} (\bibinfo {year} {2017})}\BibitemShut {NoStop}%
\bibitem [{\citenamefont {Mondaini}\ and\ \citenamefont
  {Cai}(2017)}]{mondaini2017many}%
  \BibitemOpen
  \bibfield  {author} {\bibinfo {author} {\bibfnamefont {R.}~\bibnamefont
  {Mondaini}}\ and\ \bibinfo {author} {\bibfnamefont {Z.}~\bibnamefont {Cai}},\
  }\bibfield  {title} {\bibinfo {title} {Many-body self-localization in a
  translation-invariant {Hamiltonian}},\ }\href
  {https://doi.org/10.1103/PhysRevB.96.035153} {\bibfield  {journal} {\bibinfo
  {journal} {Phys. Rev. B}\ }\textbf {\bibinfo {volume} {96}},\ \bibinfo
  {pages} {035153} (\bibinfo {year} {2017})}\BibitemShut {NoStop}%
\bibitem [{\citenamefont {Bols}\ and\ \citenamefont
  {De~Roeck}(2018)}]{bols2018asymptotic}%
  \BibitemOpen
  \bibfield  {author} {\bibinfo {author} {\bibfnamefont {A.}~\bibnamefont
  {Bols}}\ and\ \bibinfo {author} {\bibfnamefont {W.}~\bibnamefont
  {De~Roeck}},\ }\bibfield  {title} {\bibinfo {title} {Asymptotic localization
  in the {Bose}-{Hubbard} model},\ }\href {https://doi.org/10.1063/1.5022757}
  {\bibfield  {journal} {\bibinfo  {journal} {J. Math. Phys.}\ }\textbf
  {\bibinfo {volume} {59}},\ \bibinfo {pages} {021901} (\bibinfo {year}
  {2018})}\BibitemShut {NoStop}%
\bibitem [{\citenamefont {Michailidis}\ \emph {et~al.}(2018)\citenamefont
  {Michailidis}, \citenamefont {Žnidarič}, \citenamefont {Medvedyeva},
  \citenamefont {Abanin}, \citenamefont {Prosen},\ and\ \citenamefont
  {Papić}}]{michailidis2018slow}%
  \BibitemOpen
  \bibfield  {author} {\bibinfo {author} {\bibfnamefont {A.~A.}\ \bibnamefont
  {Michailidis}}, \bibinfo {author} {\bibfnamefont {M.}~\bibnamefont
  {Žnidarič}}, \bibinfo {author} {\bibfnamefont {M.}~\bibnamefont
  {Medvedyeva}}, \bibinfo {author} {\bibfnamefont {D.~A.}\ \bibnamefont
  {Abanin}}, \bibinfo {author} {\bibfnamefont {T.}~\bibnamefont {Prosen}},\
  and\ \bibinfo {author} {\bibfnamefont {Z.}~\bibnamefont {Papić}},\
  }\bibfield  {title} {\bibinfo {title} {Slow dynamics in translation-invariant
  quantum lattice models},\ }\href {https://doi.org/10.1103/PhysRevB.97.104307}
  {\bibfield  {journal} {\bibinfo  {journal} {Phys. Rev. B}\ }\textbf {\bibinfo
  {volume} {97}},\ \bibinfo {pages} {104307} (\bibinfo {year}
  {2018})}\BibitemShut {NoStop}%
\bibitem [{\citenamefont {Li}\ \emph {et~al.}(2021)\citenamefont {Li},
  \citenamefont {Deng},\ and\ \citenamefont {Santos}}]{li2021hilbert}%
  \BibitemOpen
  \bibfield  {author} {\bibinfo {author} {\bibfnamefont {W.-H.}\ \bibnamefont
  {Li}}, \bibinfo {author} {\bibfnamefont {X.}~\bibnamefont {Deng}},\ and\
  \bibinfo {author} {\bibfnamefont {L.}~\bibnamefont {Santos}},\ }\bibfield
  {title} {\bibinfo {title} {Hilbert {Space} {Shattering} and {Disorder}-{Free}
  {Localization} in {Polar} {Lattice} {Gases}},\ }\href
  {https://doi.org/10.1103/PhysRevLett.127.260601} {\bibfield  {journal}
  {\bibinfo  {journal} {Phys. Rev. Lett.}\ }\textbf {\bibinfo {volume} {127}},\
  \bibinfo {pages} {260601} (\bibinfo {year} {2021})}\BibitemShut {NoStop}%
\bibitem [{\citenamefont {Spielman}\ \emph {et~al.}(2024)\citenamefont
  {Spielman}, \citenamefont {Handian}, \citenamefont {Inman}, \citenamefont
  {Carroll},\ and\ \citenamefont {Noel}}]{spielman2024quantum}%
  \BibitemOpen
  \bibfield  {author} {\bibinfo {author} {\bibfnamefont {S.~E.}\ \bibnamefont
  {Spielman}}, \bibinfo {author} {\bibfnamefont {A.}~\bibnamefont {Handian}},
  \bibinfo {author} {\bibfnamefont {N.~P.}\ \bibnamefont {Inman}}, \bibinfo
  {author} {\bibfnamefont {T.~J.}\ \bibnamefont {Carroll}},\ and\ \bibinfo
  {author} {\bibfnamefont {M.~W.}\ \bibnamefont {Noel}},\ }\bibfield  {title}
  {\bibinfo {title} {Quantum many-body scars in few-body dipole-dipole
  interactions},\ }\href {https://doi.org/10.1103/PhysRevResearch.6.043086}
  {\bibfield  {journal} {\bibinfo  {journal} {Phys. Rev. Res.}\ }\textbf
  {\bibinfo {volume} {6}},\ \bibinfo {pages} {043086} (\bibinfo {year}
  {2024})}\BibitemShut {NoStop}%
\bibitem [{\citenamefont {Turner}\ \emph
  {et~al.}(2018{\natexlab{a}})\citenamefont {Turner}, \citenamefont
  {Michailidis}, \citenamefont {Abanin}, \citenamefont {Serbyn},\ and\
  \citenamefont {Papić}}]{turner2018weak}%
  \BibitemOpen
  \bibfield  {author} {\bibinfo {author} {\bibfnamefont {C.~J.}\ \bibnamefont
  {Turner}}, \bibinfo {author} {\bibfnamefont {A.~A.}\ \bibnamefont
  {Michailidis}}, \bibinfo {author} {\bibfnamefont {D.~A.}\ \bibnamefont
  {Abanin}}, \bibinfo {author} {\bibfnamefont {M.}~\bibnamefont {Serbyn}},\
  and\ \bibinfo {author} {\bibfnamefont {Z.}~\bibnamefont {Papić}},\
  }\bibfield  {title} {\bibinfo {title} {Weak ergodicity breaking from quantum
  many-body scars},\ }\href {https://doi.org/10.1038/s41567-018-0137-5}
  {\bibfield  {journal} {\bibinfo  {journal} {Nature Phys}\ }\textbf {\bibinfo
  {volume} {14}},\ \bibinfo {pages} {745} (\bibinfo {year}
  {2018}{\natexlab{a}})}\BibitemShut {NoStop}%
\bibitem [{\citenamefont {van Horssen}\ \emph {et~al.}(2015)\citenamefont {van
  Horssen}, \citenamefont {Levi},\ and\ \citenamefont
  {Garrahan}}]{horssen2015dynamics}%
  \BibitemOpen
  \bibfield  {author} {\bibinfo {author} {\bibfnamefont {M.}~\bibnamefont {van
  Horssen}}, \bibinfo {author} {\bibfnamefont {E.}~\bibnamefont {Levi}},\ and\
  \bibinfo {author} {\bibfnamefont {J.~P.}\ \bibnamefont {Garrahan}},\
  }\bibfield  {title} {\bibinfo {title} {Dynamics of many-body localization in
  a translation-invariant quantum glass model},\ }\href
  {https://doi.org/10.1103/PhysRevB.92.100305} {\bibfield  {journal} {\bibinfo
  {journal} {Phys. Rev. B}\ }\textbf {\bibinfo {volume} {92}},\ \bibinfo
  {pages} {100305} (\bibinfo {year} {2015})}\BibitemShut {NoStop}%
\bibitem [{\citenamefont {Moudgalya}\ \emph {et~al.}(2020)\citenamefont
  {Moudgalya}, \citenamefont {Prem}, \citenamefont {Nandkishore}, \citenamefont
  {Regnault},\ and\ \citenamefont {Bernevig}}]{moudgalya2020thermalization}%
  \BibitemOpen
  \bibfield  {author} {\bibinfo {author} {\bibfnamefont {S.}~\bibnamefont
  {Moudgalya}}, \bibinfo {author} {\bibfnamefont {A.}~\bibnamefont {Prem}},
  \bibinfo {author} {\bibfnamefont {R.}~\bibnamefont {Nandkishore}}, \bibinfo
  {author} {\bibfnamefont {N.}~\bibnamefont {Regnault}},\ and\ \bibinfo
  {author} {\bibfnamefont {B.~A.}\ \bibnamefont {Bernevig}},\ }\bibfield
  {title} {\bibinfo {title} {Thermalization and {Its} {Absence} within {Krylov}
  {Subspaces} of a {Constrained} {Hamiltonian}},\ }in\ \href
  {https://doi.org/10.1142/9789811231711_0009} {\emph {\bibinfo {booktitle}
  {Memorial {Volume} for {Shoucheng} {Zhang}}}}\ (\bibinfo  {publisher} {WORLD
  SCIENTIFIC},\ \bibinfo {year} {2020})\ pp.\ \bibinfo {pages}
  {147--209}\BibitemShut {NoStop}%
\bibitem [{\citenamefont {Brighi}\ \emph {et~al.}(2023)\citenamefont {Brighi},
  \citenamefont {Ljubotina},\ and\ \citenamefont {Serbyn}}]{brighi2023hilbert}%
  \BibitemOpen
  \bibfield  {author} {\bibinfo {author} {\bibfnamefont {P.}~\bibnamefont
  {Brighi}}, \bibinfo {author} {\bibfnamefont {M.}~\bibnamefont {Ljubotina}},\
  and\ \bibinfo {author} {\bibfnamefont {M.}~\bibnamefont {Serbyn}},\
  }\bibfield  {title} {\bibinfo {title} {Hilbert space fragmentation and slow
  dynamics in particle-conserving quantum {East} models},\ }\href
  {https://doi.org/10.21468/SciPostPhys.15.3.093} {\bibfield  {journal}
  {\bibinfo  {journal} {SciPost Physics}\ }\textbf {\bibinfo {volume} {15}},\
  \bibinfo {pages} {093} (\bibinfo {year} {2023})}\BibitemShut {NoStop}%
\bibitem [{\citenamefont {Yang}\ \emph {et~al.}(2020)\citenamefont {Yang},
  \citenamefont {Liu}, \citenamefont {Gorshkov},\ and\ \citenamefont
  {Iadecola}}]{yang2020hilbert}%
  \BibitemOpen
  \bibfield  {author} {\bibinfo {author} {\bibfnamefont {Z.-C.}\ \bibnamefont
  {Yang}}, \bibinfo {author} {\bibfnamefont {F.}~\bibnamefont {Liu}}, \bibinfo
  {author} {\bibfnamefont {A.~V.}\ \bibnamefont {Gorshkov}},\ and\ \bibinfo
  {author} {\bibfnamefont {T.}~\bibnamefont {Iadecola}},\ }\bibfield  {title}
  {\bibinfo {title} {Hilbert-{Space} {Fragmentation} from {Strict}
  {Confinement}},\ }\href {https://doi.org/10.1103/PhysRevLett.124.207602}
  {\bibfield  {journal} {\bibinfo  {journal} {Phys. Rev. Lett.}\ }\textbf
  {\bibinfo {volume} {124}},\ \bibinfo {pages} {207602} (\bibinfo {year}
  {2020})}\BibitemShut {NoStop}%
\bibitem [{\citenamefont {Yang}(2022)}]{yang2022distinction}%
  \BibitemOpen
  \bibfield  {author} {\bibinfo {author} {\bibfnamefont {Z.-C.}\ \bibnamefont
  {Yang}},\ }\bibfield  {title} {\bibinfo {title} {Distinction between
  transport and {R}{\textbackslash}'enyi entropy growth in kinetically
  constrained models},\ }\href {https://doi.org/10.1103/PhysRevB.106.L220303}
  {\bibfield  {journal} {\bibinfo  {journal} {Phys. Rev. B}\ }\textbf {\bibinfo
  {volume} {106}},\ \bibinfo {pages} {L220303} (\bibinfo {year}
  {2022})}\BibitemShut {NoStop}%
\bibitem [{\citenamefont {Causer}\ \emph {et~al.}(2024)\citenamefont {Causer},
  \citenamefont {Bañuls},\ and\ \citenamefont
  {Garrahan}}]{causer2024nonthermal}%
  \BibitemOpen
  \bibfield  {author} {\bibinfo {author} {\bibfnamefont {L.}~\bibnamefont
  {Causer}}, \bibinfo {author} {\bibfnamefont {M.~C.}\ \bibnamefont
  {Bañuls}},\ and\ \bibinfo {author} {\bibfnamefont {J.~P.}\ \bibnamefont
  {Garrahan}},\ }\bibfield  {title} {\bibinfo {title} {Nonthermal eigenstates
  and slow relaxation in quantum {Fredkin} spin chains},\ }\href
  {https://doi.org/10.1103/PhysRevB.110.134322} {\bibfield  {journal} {\bibinfo
   {journal} {Phys. Rev. B}\ }\textbf {\bibinfo {volume} {110}},\ \bibinfo
  {pages} {134322} (\bibinfo {year} {2024})}\BibitemShut {NoStop}%
\bibitem [{\citenamefont {Dias}(2000)}]{dias2000exact}%
  \BibitemOpen
  \bibfield  {author} {\bibinfo {author} {\bibfnamefont {R.~G.}\ \bibnamefont
  {Dias}},\ }\bibfield  {title} {\bibinfo {title} {Exact solution of the strong
  coupling \$t{\textbackslash}ensuremath\{-\}{V}\$ model with twisted boundary
  conditions},\ }\href {https://doi.org/10.1103/PhysRevB.62.7791} {\bibfield
  {journal} {\bibinfo  {journal} {Phys. Rev. B}\ }\textbf {\bibinfo {volume}
  {62}},\ \bibinfo {pages} {7791} (\bibinfo {year} {2000})}\BibitemShut
  {NoStop}%
\bibitem [{\citenamefont {Zadnik}\ and\ \citenamefont
  {Fagotti}(2021)}]{zadnik2021folded}%
  \BibitemOpen
  \bibfield  {author} {\bibinfo {author} {\bibfnamefont {L.}~\bibnamefont
  {Zadnik}}\ and\ \bibinfo {author} {\bibfnamefont {M.}~\bibnamefont
  {Fagotti}},\ }\bibfield  {title} {\bibinfo {title} {The {Folded} {Spin}-1/2
  {XXZ} {Model}: {I}. {Diagonalisation}, {Jamming}, and {Ground} {State}
  {Properties}},\ }\href {https://doi.org/10.21468/SciPostPhysCore.4.2.010}
  {\bibfield  {journal} {\bibinfo  {journal} {SciPost Physics Core}\ }\textbf
  {\bibinfo {volume} {4}},\ \bibinfo {pages} {010} (\bibinfo {year}
  {2021})}\BibitemShut {NoStop}%
\bibitem [{\citenamefont {Zadnik}\ \emph {et~al.}(2021)\citenamefont {Zadnik},
  \citenamefont {Bidzhiev},\ and\ \citenamefont {Fagotti}}]{zadnik2021foldeda}%
  \BibitemOpen
  \bibfield  {author} {\bibinfo {author} {\bibfnamefont {L.}~\bibnamefont
  {Zadnik}}, \bibinfo {author} {\bibfnamefont {K.}~\bibnamefont {Bidzhiev}},\
  and\ \bibinfo {author} {\bibfnamefont {M.}~\bibnamefont {Fagotti}},\
  }\bibfield  {title} {\bibinfo {title} {The folded spin-1/2 {XXZ} model: {II}.
  {Thermodynamics} and hydrodynamics with a minimal set of charges},\ }\href
  {https://doi.org/10.21468/SciPostPhys.10.5.099} {\bibfield  {journal}
  {\bibinfo  {journal} {SciPost Physics}\ }\textbf {\bibinfo {volume} {10}},\
  \bibinfo {pages} {099} (\bibinfo {year} {2021})}\BibitemShut {NoStop}%
\bibitem [{\citenamefont {Khemani}\ \emph {et~al.}(2020)\citenamefont
  {Khemani}, \citenamefont {Hermele},\ and\ \citenamefont
  {Nandkishore}}]{khemani2020localization}%
  \BibitemOpen
  \bibfield  {author} {\bibinfo {author} {\bibfnamefont {V.}~\bibnamefont
  {Khemani}}, \bibinfo {author} {\bibfnamefont {M.}~\bibnamefont {Hermele}},\
  and\ \bibinfo {author} {\bibfnamefont {R.}~\bibnamefont {Nandkishore}},\
  }\bibfield  {title} {\bibinfo {title} {Localization from {Hilbert} space
  shattering: {From} theory to physical realizations},\ }\href
  {https://doi.org/10.1103/PhysRevB.101.174204} {\bibfield  {journal} {\bibinfo
   {journal} {Phys. Rev. B}\ }\textbf {\bibinfo {volume} {101}},\ \bibinfo
  {pages} {174204} (\bibinfo {year} {2020})}\BibitemShut {NoStop}%
\bibitem [{\citenamefont {Sala}\ \emph {et~al.}(2020)\citenamefont {Sala},
  \citenamefont {Rakovszky}, \citenamefont {Verresen}, \citenamefont {Knap},\
  and\ \citenamefont {Pollmann}}]{sala2020ergodicity}%
  \BibitemOpen
  \bibfield  {author} {\bibinfo {author} {\bibfnamefont {P.}~\bibnamefont
  {Sala}}, \bibinfo {author} {\bibfnamefont {T.}~\bibnamefont {Rakovszky}},
  \bibinfo {author} {\bibfnamefont {R.}~\bibnamefont {Verresen}}, \bibinfo
  {author} {\bibfnamefont {M.}~\bibnamefont {Knap}},\ and\ \bibinfo {author}
  {\bibfnamefont {F.}~\bibnamefont {Pollmann}},\ }\bibfield  {title} {\bibinfo
  {title} {Ergodicity {Breaking} {Arising} from {Hilbert} {Space}
  {Fragmentation} in {Dipole}-{Conserving} {Hamiltonians}},\ }\href
  {https://doi.org/10.1103/PhysRevX.10.011047} {\bibfield  {journal} {\bibinfo
  {journal} {Phys. Rev. X}\ }\textbf {\bibinfo {volume} {10}},\ \bibinfo
  {pages} {011047} (\bibinfo {year} {2020})}\BibitemShut {NoStop}%
\bibitem [{\citenamefont {Turner}\ \emph
  {et~al.}(2018{\natexlab{b}})\citenamefont {Turner}, \citenamefont
  {Michailidis}, \citenamefont {Abanin}, \citenamefont {Serbyn},\ and\
  \citenamefont {Papić}}]{turner2018quantum}%
  \BibitemOpen
  \bibfield  {author} {\bibinfo {author} {\bibfnamefont {C.~J.}\ \bibnamefont
  {Turner}}, \bibinfo {author} {\bibfnamefont {A.~A.}\ \bibnamefont
  {Michailidis}}, \bibinfo {author} {\bibfnamefont {D.~A.}\ \bibnamefont
  {Abanin}}, \bibinfo {author} {\bibfnamefont {M.}~\bibnamefont {Serbyn}},\
  and\ \bibinfo {author} {\bibfnamefont {Z.}~\bibnamefont {Papić}},\
  }\bibfield  {title} {\bibinfo {title} {Quantum scarred eigenstates in a
  {Rydberg} atom chain: {Entanglement}, breakdown of thermalization, and
  stability to perturbations},\ }\href
  {https://doi.org/10.1103/PhysRevB.98.155134} {\bibfield  {journal} {\bibinfo
  {journal} {Phys. Rev. B}\ }\textbf {\bibinfo {volume} {98}},\ \bibinfo
  {pages} {155134} (\bibinfo {year} {2018}{\natexlab{b}})}\BibitemShut
  {NoStop}%
\bibitem [{\citenamefont {Serbyn}\ \emph {et~al.}(2021)\citenamefont {Serbyn},
  \citenamefont {Abanin},\ and\ \citenamefont {Papić}}]{serbyn2021quantum}%
  \BibitemOpen
  \bibfield  {author} {\bibinfo {author} {\bibfnamefont {M.}~\bibnamefont
  {Serbyn}}, \bibinfo {author} {\bibfnamefont {D.~A.}\ \bibnamefont {Abanin}},\
  and\ \bibinfo {author} {\bibfnamefont {Z.}~\bibnamefont {Papić}},\
  }\bibfield  {title} {\bibinfo {title} {Quantum many-body scars and weak
  breaking of ergodicity},\ }\href {https://doi.org/10.1038/s41567-021-01230-2}
  {\bibfield  {journal} {\bibinfo  {journal} {Nat. Phys.}\ }\textbf {\bibinfo
  {volume} {17}},\ \bibinfo {pages} {675} (\bibinfo {year} {2021})}\BibitemShut
  {NoStop}%
\bibitem [{\citenamefont {Sánchez}\ \emph {et~al.}(2018)\citenamefont
  {Sánchez}, \citenamefont {Varma},\ and\ \citenamefont
  {Oganesyan}}]{sanchez2018anomalous}%
  \BibitemOpen
  \bibfield  {author} {\bibinfo {author} {\bibfnamefont {R.~J.}\ \bibnamefont
  {Sánchez}}, \bibinfo {author} {\bibfnamefont {V.~K.}\ \bibnamefont
  {Varma}},\ and\ \bibinfo {author} {\bibfnamefont {V.}~\bibnamefont
  {Oganesyan}},\ }\bibfield  {title} {\bibinfo {title} {Anomalous and regular
  transport in spin-\${\textbackslash}frac\{1\}\{2\}\$ chains: ac
  conductivity},\ }\href {https://doi.org/10.1103/PhysRevB.98.054415}
  {\bibfield  {journal} {\bibinfo  {journal} {Phys. Rev. B}\ }\textbf {\bibinfo
  {volume} {98}},\ \bibinfo {pages} {054415} (\bibinfo {year}
  {2018})}\BibitemShut {NoStop}%
\bibitem [{\citenamefont {Singh}\ \emph {et~al.}(2021)\citenamefont {Singh},
  \citenamefont {Ware}, \citenamefont {Vasseur},\ and\ \citenamefont
  {Friedman}}]{singh2021subdiffusion}%
  \BibitemOpen
  \bibfield  {author} {\bibinfo {author} {\bibfnamefont {H.}~\bibnamefont
  {Singh}}, \bibinfo {author} {\bibfnamefont {B.}~\bibnamefont {Ware}},
  \bibinfo {author} {\bibfnamefont {R.}~\bibnamefont {Vasseur}},\ and\ \bibinfo
  {author} {\bibfnamefont {A.~J.}\ \bibnamefont {Friedman}},\ }\bibfield
  {title} {\bibinfo {title} {Subdiffusion and many-body quantum chaos with
  kinetic constraints},\ }\href
  {https://doi.org/10.1103/PhysRevLett.127.230602} {\bibfield  {journal}
  {\bibinfo  {journal} {Phys. Rev. Lett.}\ }\textbf {\bibinfo {volume} {127}},\
  \bibinfo {pages} {230602} (\bibinfo {year} {2021})}\BibitemShut {NoStop}%
\bibitem [{\citenamefont {De~Nardis}\ \emph {et~al.}(2022)\citenamefont
  {De~Nardis}, \citenamefont {Gopalakrishnan}, \citenamefont {Vasseur},\ and\
  \citenamefont {Ware}}]{nardis2022subdiffusive}%
  \BibitemOpen
  \bibfield  {author} {\bibinfo {author} {\bibfnamefont {J.}~\bibnamefont
  {De~Nardis}}, \bibinfo {author} {\bibfnamefont {S.}~\bibnamefont
  {Gopalakrishnan}}, \bibinfo {author} {\bibfnamefont {R.}~\bibnamefont
  {Vasseur}},\ and\ \bibinfo {author} {\bibfnamefont {B.}~\bibnamefont
  {Ware}},\ }\bibfield  {title} {\bibinfo {title} {Subdiffusive hydrodynamics
  of nearly integrable anisotropic spin chains},\ }\href
  {https://doi.org/10.1073/pnas.2202823119} {\bibfield  {journal} {\bibinfo
  {journal} {Proc. Natl. Acad. Sci. U.S.A.}\ }\textbf {\bibinfo {volume}
  {119}},\ \bibinfo {pages} {e2202823119} (\bibinfo {year} {2022})}\BibitemShut
  {NoStop}%
\bibitem [{\citenamefont {Chen}\ \emph {et~al.}(2024)\citenamefont {Chen},
  \citenamefont {Chen},\ and\ \citenamefont {Wang}}]{chen2024superdiffusive}%
  \BibitemOpen
  \bibfield  {author} {\bibinfo {author} {\bibfnamefont {C.}~\bibnamefont
  {Chen}}, \bibinfo {author} {\bibfnamefont {Y.}~\bibnamefont {Chen}},\ and\
  \bibinfo {author} {\bibfnamefont {X.}~\bibnamefont {Wang}},\ }\bibfield
  {title} {\bibinfo {title} {Superdiffusive to ballistic transport in
  nonintegrable {Rydberg} simulator},\ }\href
  {https://doi.org/10.1038/s41534-024-00884-z} {\bibfield  {journal} {\bibinfo
  {journal} {npj Quantum Inf}\ }\textbf {\bibinfo {volume} {10}},\ \bibinfo
  {pages} {1} (\bibinfo {year} {2024})}\BibitemShut {NoStop}%
\bibitem [{\citenamefont {Mott}\ and\ \citenamefont
  {Davis}(1968)}]{mott1968conduction}%
  \BibitemOpen
  \bibfield  {author} {\bibinfo {author} {\bibfnamefont {N.~F.}\ \bibnamefont
  {Mott}}\ and\ \bibinfo {author} {\bibfnamefont {E.~A.}\ \bibnamefont
  {Davis}},\ }\bibfield  {title} {\bibinfo {title} {Conduction in
  non-crystalline systems},\ }\href {https://doi.org/10.1080/14786436808223201}
  {\bibfield  {journal} {\bibinfo  {journal} {Philosophical Magazine}\ }\textbf
  {\bibinfo {volume} {17}},\ \bibinfo {pages} {1269} (\bibinfo {year}
  {1968})}\BibitemShut {NoStop}%
\bibitem [{\citenamefont {Gopalakrishnan}\ \emph {et~al.}(2015)\citenamefont
  {Gopalakrishnan}, \citenamefont {Müller}, \citenamefont {Khemani},
  \citenamefont {Knap}, \citenamefont {Demler},\ and\ \citenamefont
  {Huse}}]{gopalakrishnan2015low}%
  \BibitemOpen
  \bibfield  {author} {\bibinfo {author} {\bibfnamefont {S.}~\bibnamefont
  {Gopalakrishnan}}, \bibinfo {author} {\bibfnamefont {M.}~\bibnamefont
  {Müller}}, \bibinfo {author} {\bibfnamefont {V.}~\bibnamefont {Khemani}},
  \bibinfo {author} {\bibfnamefont {M.}~\bibnamefont {Knap}}, \bibinfo {author}
  {\bibfnamefont {E.}~\bibnamefont {Demler}},\ and\ \bibinfo {author}
  {\bibfnamefont {D.~A.}\ \bibnamefont {Huse}},\ }\bibfield  {title} {\bibinfo
  {title} {Low-frequency conductivity in many-body localized systems},\ }\href
  {https://doi.org/10.1103/PhysRevB.92.104202} {\bibfield  {journal} {\bibinfo
  {journal} {Phys. Rev. B}\ }\textbf {\bibinfo {volume} {92}},\ \bibinfo
  {pages} {104202} (\bibinfo {year} {2015})}\BibitemShut {NoStop}%
\bibitem [{\citenamefont {Giamarchi}(2004)}]{giamarchi2004quantum}%
  \BibitemOpen
  \bibfield  {author} {\bibinfo {author} {\bibfnamefont {T.}~\bibnamefont
  {Giamarchi}},\ }\href@noop {} {\emph {\bibinfo {title} {Quantum physics in
  one dimension}}},\ Vol.\ \bibinfo {volume} {121}\ (\bibinfo  {publisher}
  {Oxford university press},\ \bibinfo {year} {2004})\BibitemShut {NoStop}%
\bibitem [{Note1()}]{Note1}%
  \BibitemOpen
  \bibinfo {note} {At infinite $\Delta $, these classical spin configurations
  are exact eigenstates. So they are used to denote each bound states. At
  finite $\Delta $, they are approximately eigenstates, with the amplitudes of
  tunneling to other configurations at the order of $O(1/\Delta )$. Therefore,
  for the latter case (especially at large $\Delta $), it is still legible to
  use the bit-strings to denote specific types of bound states.}\BibitemShut
  {Stop}%
\bibitem [{\citenamefont {Steinigeweg}\ \emph {et~al.}(2014)\citenamefont
  {Steinigeweg}, \citenamefont {Gemmer},\ and\ \citenamefont
  {Brenig}}]{steinigeweg2014spin}%
  \BibitemOpen
  \bibfield  {author} {\bibinfo {author} {\bibfnamefont {R.}~\bibnamefont
  {Steinigeweg}}, \bibinfo {author} {\bibfnamefont {J.}~\bibnamefont
  {Gemmer}},\ and\ \bibinfo {author} {\bibfnamefont {W.}~\bibnamefont
  {Brenig}},\ }\bibfield  {title} {\bibinfo {title} {Spin-{Current}
  {Autocorrelations} from {Single} {Pure}-{State} {Propagation}},\ }\href
  {https://doi.org/10.1103/PhysRevLett.112.120601} {\bibfield  {journal}
  {\bibinfo  {journal} {Phys. Rev. Lett.}\ }\textbf {\bibinfo {volume} {112}},\
  \bibinfo {pages} {120601} (\bibinfo {year} {2014})}\BibitemShut {NoStop}%
\bibitem [{\citenamefont {Steinigeweg}\ \emph
  {et~al.}(2015{\natexlab{a}})\citenamefont {Steinigeweg}, \citenamefont
  {Gemmer},\ and\ \citenamefont {Brenig}}]{steinigeweg2015spin}%
  \BibitemOpen
  \bibfield  {author} {\bibinfo {author} {\bibfnamefont {R.}~\bibnamefont
  {Steinigeweg}}, \bibinfo {author} {\bibfnamefont {J.}~\bibnamefont
  {Gemmer}},\ and\ \bibinfo {author} {\bibfnamefont {W.}~\bibnamefont
  {Brenig}},\ }\bibfield  {title} {\bibinfo {title} {Spin and energy currents
  in integrable and nonintegrable spin-\${\textbackslash}frac\{1\}\{2\}\$
  chains: {A} typicality approach to real-time autocorrelations},\ }\href
  {https://doi.org/10.1103/PhysRevB.91.104404} {\bibfield  {journal} {\bibinfo
  {journal} {Phys. Rev. B}\ }\textbf {\bibinfo {volume} {91}},\ \bibinfo
  {pages} {104404} (\bibinfo {year} {2015}{\natexlab{a}})}\BibitemShut
  {NoStop}%
\bibitem [{Note2()}]{Note2}%
  \BibitemOpen
  \bibinfo {note} {For the effective Hamiltonian, the spin current operator is
  rewritten as $\protect \hat J = J\DOTSB \sum@ \slimits@ _i {P_{r,i}(S_i^xS_{i
  + 1}^y - S_i^yS_{i + 1}^x)}$.}\BibitemShut {Stop}%
\bibitem [{\citenamefont {Vlijm}\ \emph {et~al.}(2015)\citenamefont {Vlijm},
  \citenamefont {Ganahl}, \citenamefont {Fioretto}, \citenamefont {Brockmann},
  \citenamefont {Haque}, \citenamefont {Evertz},\ and\ \citenamefont
  {Caux}}]{vlijm2015quasi}%
  \BibitemOpen
  \bibfield  {author} {\bibinfo {author} {\bibfnamefont {R.}~\bibnamefont
  {Vlijm}}, \bibinfo {author} {\bibfnamefont {M.}~\bibnamefont {Ganahl}},
  \bibinfo {author} {\bibfnamefont {D.}~\bibnamefont {Fioretto}}, \bibinfo
  {author} {\bibfnamefont {M.}~\bibnamefont {Brockmann}}, \bibinfo {author}
  {\bibfnamefont {M.}~\bibnamefont {Haque}}, \bibinfo {author} {\bibfnamefont
  {H.~G.}\ \bibnamefont {Evertz}},\ and\ \bibinfo {author} {\bibfnamefont
  {J.-S.}\ \bibnamefont {Caux}},\ }\bibfield  {title} {\bibinfo {title}
  {Quasi-soliton scattering in quantum spin chains},\ }\href
  {https://doi.org/10.1103/PhysRevB.92.214427} {\bibfield  {journal} {\bibinfo
  {journal} {Phys. Rev. B}\ }\textbf {\bibinfo {volume} {92}},\ \bibinfo
  {pages} {214427} (\bibinfo {year} {2015})}\BibitemShut {NoStop}%
\bibitem [{\citenamefont {Gopalakrishnan}\ and\ \citenamefont
  {Vasseur}(2019)}]{gopalakrishnan2019kinetic}%
  \BibitemOpen
  \bibfield  {author} {\bibinfo {author} {\bibfnamefont {S.}~\bibnamefont
  {Gopalakrishnan}}\ and\ \bibinfo {author} {\bibfnamefont {R.}~\bibnamefont
  {Vasseur}},\ }\bibfield  {title} {\bibinfo {title} {Kinetic {Theory} of
  {Spin} {Diffusion} and {Superdiffusion} in \${XXZ}\$ {Spin} {Chains}},\
  }\href {https://doi.org/10.1103/PhysRevLett.122.127202} {\bibfield  {journal}
  {\bibinfo  {journal} {Phys. Rev. Lett.}\ }\textbf {\bibinfo {volume} {122}},\
  \bibinfo {pages} {127202} (\bibinfo {year} {2019})}\BibitemShut {NoStop}%
\bibitem [{\citenamefont {Rigol}\ and\ \citenamefont
  {Shastry}(2008)}]{rigol2008drude}%
  \BibitemOpen
  \bibfield  {author} {\bibinfo {author} {\bibfnamefont {M.}~\bibnamefont
  {Rigol}}\ and\ \bibinfo {author} {\bibfnamefont {B.~S.}\ \bibnamefont
  {Shastry}},\ }\bibfield  {title} {\bibinfo {title} {Drude weight in systems
  with open boundary conditions},\ }\href
  {https://doi.org/10.1103/PhysRevB.77.161101} {\bibfield  {journal} {\bibinfo
  {journal} {Phys. Rev. B}\ }\textbf {\bibinfo {volume} {77}},\ \bibinfo
  {pages} {161101} (\bibinfo {year} {2008})}\BibitemShut {NoStop}%
\bibitem [{\citenamefont {Spickerman}(1982)}]{spickerman1982binet}%
  \BibitemOpen
  \bibfield  {author} {\bibinfo {author} {\bibfnamefont {W.~R.}\ \bibnamefont
  {Spickerman}},\ }\bibfield  {title} {\bibinfo {title} {Binet’s formula for
  the {Tribonacci} sequence},\ }\href
  {https://www.mathstat.dal.ca/FQ/Scanned/20-2/spickerman.pdf} {\bibfield
  {journal} {\bibinfo  {journal} {Fibonacci Quart}\ }\textbf {\bibinfo {volume}
  {20}},\ \bibinfo {pages} {118} (\bibinfo {year} {1982})}\BibitemShut
  {NoStop}%
\bibitem [{\citenamefont {Weinberg}\ and\ \citenamefont
  {Bukov}(2017)}]{weinberg2017quspin}%
  \BibitemOpen
  \bibfield  {author} {\bibinfo {author} {\bibfnamefont {P.}~\bibnamefont
  {Weinberg}}\ and\ \bibinfo {author} {\bibfnamefont {M.}~\bibnamefont
  {Bukov}},\ }\bibfield  {title} {\bibinfo {title} {{QuSpin}: a {Python}
  package for dynamics and exact diagonalisation of quantum many body systems
  part {I}: spin chains},\ }\href
  {https://doi.org/10.21468/SciPostPhys.2.1.003} {\bibfield  {journal}
  {\bibinfo  {journal} {SciPost Phys.}\ }\textbf {\bibinfo {volume} {2}},\
  \bibinfo {pages} {003} (\bibinfo {year} {2017})}\BibitemShut {NoStop}%
\bibitem [{\citenamefont {Weinberg}\ and\ \citenamefont
  {Bukov}(2019)}]{weinberg2019quspin}%
  \BibitemOpen
  \bibfield  {author} {\bibinfo {author} {\bibfnamefont {P.}~\bibnamefont
  {Weinberg}}\ and\ \bibinfo {author} {\bibfnamefont {M.}~\bibnamefont
  {Bukov}},\ }\bibfield  {title} {\bibinfo {title} {{QuSpin}: a {Python}
  package for dynamics and exact diagonalisation of quantum many body systems.
  {Part} {II}: bosons, fermions and higher spins},\ }\href
  {https://doi.org/10.21468/SciPostPhys.7.2.020} {\bibfield  {journal}
  {\bibinfo  {journal} {SciPost Physics}\ }\textbf {\bibinfo {volume} {7}},\
  \bibinfo {pages} {020} (\bibinfo {year} {2019})}\BibitemShut {NoStop}%
\bibitem [{\citenamefont {Steinigeweg}\ \emph
  {et~al.}(2015{\natexlab{b}})\citenamefont {Steinigeweg}, \citenamefont
  {Gemmer},\ and\ \citenamefont {Brenig}}]{steinigeweg2015spin_supp}%
  \BibitemOpen
  \bibfield  {author} {\bibinfo {author} {\bibfnamefont {R.}~\bibnamefont
  {Steinigeweg}}, \bibinfo {author} {\bibfnamefont {J.}~\bibnamefont
  {Gemmer}},\ and\ \bibinfo {author} {\bibfnamefont {W.}~\bibnamefont
  {Brenig}},\ }\bibfield  {title} {\bibinfo {title} {Spin and energy currents
  in integrable and nonintegrable spin-\${\textbackslash}frac\{1\}\{2\}\$
  chains: {A} typicality approach to real-time autocorrelations},\ }\href
  {https://doi.org/10.1103/PhysRevB.91.104404} {\bibfield  {journal} {\bibinfo
  {journal} {Phys. Rev. B}\ }\textbf {\bibinfo {volume} {91}},\ \bibinfo
  {pages} {104404} (\bibinfo {year} {2015}{\natexlab{b}})}\BibitemShut
  {NoStop}%
\bibitem [{\citenamefont {Maldague}(1977)}]{maldague1977optical}%
  \BibitemOpen
  \bibfield  {author} {\bibinfo {author} {\bibfnamefont {P.~F.}\ \bibnamefont
  {Maldague}},\ }\bibfield  {title} {\bibinfo {title} {Optical spectrum of a
  {Hubbard} chain},\ }\href {https://doi.org/10.1103/PhysRevB.16.2437}
  {\bibfield  {journal} {\bibinfo  {journal} {Phys. Rev. B}\ }\textbf {\bibinfo
  {volume} {16}},\ \bibinfo {pages} {2437} (\bibinfo {year}
  {1977})}\BibitemShut {NoStop}%
\end{thebibliography}

%\end{CJK}

\end{document}